\documentclass[12pt]{article}
\usepackage{graphicx}
\usepackage{amsmath}
\usepackage{amssymb}
\usepackage{afterpage}
\usepackage{float}
\usepackage[margin=2cm]{geometry}
\restylefloat{figure}
\restylefloat{table}
\makeatletter
\renewcommand\paragraph{\@startsection{paragraph}{3}{\z@}%
                                    {3.25ex \@plus1ex \@minus.2ex}%
                                    {-1em}%
                                    {\normalfont\normalsize\bfseries}}
\makeatother

\title{A model of sympatric speciation through assortative mating}
\author{\centerline{\normalsize Franco Bagnoli$^{1,2,3}$\thanks{Electronic address: franco.bagnoli@unifi.it} ~~and Carlo Guardiani$^{3}$\thanks{Electronic address: carlo@dma.unifi.it}} \\
\centerline{\textit{\normalsize $^1$ Dipartimento di Energetica, Universit\`a di Firenze,Via S.~Marta 3, I-50139 Firenze, Italy}}\\
\centerline{\textit{\normalsize $^2$ INFN, sez. Firenze}}\\
\centerline{\textit{\normalsize $^3$ Centro Interdipartimentale per lo Studio delle Dinamiche Complesse, }} \\
\centerline{\textit{\normalsize Universit\`a di Firenze, Via Sansone 1, I-50019, Sesto Fiorentino, Italy}}
}

\date{\today}

\begin{document}

\maketitle

\begin{abstract}
A microscopic model is developed, within the frame of the theory of quantitative traits, to study the combined effect of competition and assortativity on the sympatric speciation process, \emph{i.e.} speciation in
the absence of geographical barriers. Two components of fitness are considered: a static one that describes
adaptation to environmental factors not related to the population itself, and a dynamic one that accounts for
interactions between organisms, \emph{e.g.} competition. 
A simulated annealing technique was applied in order to speed up simulations.
The simulations show that both in the case of flat and
steep static fitness landscapes, competition and assortativity do exert a synergistic effect on speciation. We
also show that competition acts as a stabilizing force against  extinction due to random sampling in a finite population. Finally, evidence is shown that speciation can be seen as a phase transition. 

\end{abstract}

\section{The problem.}

The notion of \emph{speciation} in biology refers to the splitting of an original species into two fertile, yet reproductively isolated strains. The \emph{allopatric theory}, which is currently accepted by  the
majority of biologists, claims that a  geographic  barrier  is needed
in order to break the gene flow  so as to allow two strains to evolve
a complete reproductive isolation. On the other hand, many evidences and experimental data have been
reported in recent years  strongly suggesting the possibility of a
\emph{sympatric} mechanism of speciation. For example, the
comparison of mythocondrial DNA sequences of cytochrome b performed
by Schlieven and others~\cite{ciclidi}, showed the monophyletic
origin of cichlid species living in some volcanic lakes of western
Africa. The main features of these lakes are the environmental
homogeneity and the absence of microgeographical barriers. It is thus
possible that the present diversity is the result of several events
of sympatric speciation. An increasing number of studies referring both to animal and plant species lend further support to this hypothesis~\cite{stickleback1,stickleback2,stickleback3,stickleback4,snails,lizards,Senecio,Travisano}.

The key element for sympatric speciation is \emph{assortative mating} that is, mating must be allowed only between individuals whose phenotypic distance does not exceed a given threshold. In fact, consider a population characterized by a bimodal distribution for an ecological character determining adaptation to the environment: in a regime of random mating the crossings between individuals of the two humps will produce intermediate phenotypes so that the distribution will never split. Two interesting theories have been developed to explain the evolution of assortativity. In Kondrashov and Kondrashov's theory~\cite{KK} \emph{disruptive selection} (for instance determined by a bimodal resource distribution) splits the population in two distinct ecological types that are later stabilized by the evolution of assortative mating. The theory of \emph{evolutionary branching} developed by Doebeli and Dieckmann~\cite{DD2} is more general in that it does not require disruptive selection: the population first converges in phenotype space to an  attracting fitness minimum (as a result of common ecological interactions such as competition, predation and mutualism) and then it splits into diverging phenotypic clusters. For example~\cite{DD1}, given a Gaussian resource distribution, the population first crowds on the phenotype with the highest fitness, and then, owing to the high level of competition, splits into two distinct groups that later become reproductively isolated due to selection of assortative mating.

In the present paper we will not investigate the evolution of assortativity that will be treated as a tunable parameter in order to study its interplay with competition. In particular we will show that: (1) assortativity alone is sufficient to induce speciation but one of the new species soon disappears due to random fluctuations; (2) stable species coexistence can be attained through the introduction of competition; (3) competition and assortativity do exert a synergistic effect on speciation so that high levels of assortativity can trigger speciation even in the presence of weak competition and \emph{vice versa}; (4) speciation can be thought of as a phase transition as can be deduced from the plot of variance versus competition and assortativity; (5) contrary to the traditional interpretation of Fisher's theorem, the mean fitness of the population does not always increase but it reaches a constant value (sometimes even after a decrease) as a result of the deterioration of environmental conditions, this result being consistent with Price's and Ewen's reformulation~\cite{Price,Ewens} of Fisher's theorem.  The use of a \emph{simulated annealing} method enables us to find stationary or quasi-stationary distribution in reasonably short simulation times.

In Section $2$ we describe our model and briefly outline its
implementation, providing some computational details; in Section $3$
we report the results of the simulations distinguishing between the
case of flat (subsection $3.1$) and steep (subsection $3.2$) static
fitness landscapes, while in subsection $3.3$ evidence is given of the robustness of our algorithm with respect to variations of the genome length; in Section $4$ we show that speciation can be
regarded as a phase transition; finally, in Section $5$ we draw the
conclusions of our study.  

\section{The model.}

In order to develop a microscopic evolution model,
first of all we have to establish how to represent an individual,
with the requirement of obtaining the simplest (and computationally
affordable) model still capturing the essential of phenomena under
study. 

There are many possible description levels: from the single basis to domains inside a gene to whole allele forms. Since the mutation patterns are quite complex at lower levels, we have chosen to codify the allelic forms of a gene as a discrete variable $g_i$ whose value is zero for the wild-type and then it increases according to the biological efficiencies of the resulting protein. At this level there are two main ingredients: the number of efficiency levels that are observable in a real population and the degeneracy of each level.

As a starting point, we study here the most simple choice, i.e.
a population of haploid individuals whose genome is
represented by a string $(g_1, g_2,\dots, g_L)$ 
of $L$ (binary) bits. Each bit represents a
\emph{locus} and the Boolean values it can take are regarded as
alternative allelic forms. In particular $g_i=0$  refers to the
\emph{wild-type} allele while  $g_i=1$ to the least deleterious
mutant. The phenotype $x$, in agreement with the theory of quantitative
traits, is just the sum of these bits, $x=\sum_{i=1}^L g_i$. According to this theory, in fact, quantitative characters are determined by many genes whose effects are small, similar and additive.

Even if we have studied only the simple case of two alternative allelic forms,  by analogy with the statistical mechanics of discrete (magnetic) 
systems, we do
not expect qualitative differences when a larger number of levels or
moderate degeneracy is taken into account. The presence of
epistatic interactions among genes  can indeed induce different
behaviors, but in this case we are abandoning the theory of
quantitative traits. On the other hand, a 
large degeneracy of non-wild-type levels with respect to the first one
can induce an error threshold-like transition~\cite{Bagnoli}, but this
transition needs also a relatively large difference in the phenotypic
trait corresponding to different indices, a situation which in our
opinion is not the typical one. 

A time step is composed by three subprocess: selection, recombination
and mutation.
Mutations are simply implemented by flipping a randomly chosen
element of the genome from 0 to 1 or vice versa. This kind of
mutations can only turn a phenotype $x$ into one of its neighbors
$x+1$ or $x-1$ and they are therefore referred to as \emph{short
range mutations}. The fact that both mutations $0 \rightarrow 1$ and
$1 \rightarrow 0$ occur with the same probability $\mu$ is a
coarse-grained approximation, because mutations affecting a wild-type
allele ( 0 allele) usually impair its function, but mutations on
already damaged genes (1 allele) are not very likely to restore their
activity. One should therefore expect that the frequency of the $1
\rightarrow 0$ mutations be significantly lower than that of the $0
\rightarrow 1$ mutations. The choice of equal frequencies for both
kinds of mutations, on the other hand, can be justified by assuming
that mutations are mostly due to duplications of genes or to
transposable elements that go in and out from target sites in DNA
with equal frequencies. Another limitation of our model of mutations
is that the frequency of mutation is independent of the \emph{locus}.
The frequency of mutation of a long gene, for example, should be
higher than that of a short gene, and the frequency of mutation
should be also dependent on the packing of chromatin. The
inaccuracies in our model of mutation, however, do not impair the
results of the algorithm, because, as Bagnoli and Bezzi
showed~\cite{Bagnoli2}, the occupation of fitness maxima mainly
depends on selection, while mutations only create genetic
variability. Moreover, the role of mutations in the present model is
even smaller, as genes are continuously rearranged through
recombination.     

Rigorously, in finite populations one cannot talk of true phase
transitions, and also the concept of invariant distribution is
questionable. On the other hand, the presence of random mutations
should make the system ergodic in the long time limit. 
 Unless otherwise noted, 
we have checked that the results we obtained do not depend on the size
of populations, which was typically varied from $10^3$ to $10^5$
individuals.

Reasonable values of the mutation rate would imply too long simulations in order to have independence on the initial state, especially for finite populations.
In order to speed-up simulations, we adopted two different 
strategies. The first is to use a simulated annealing technique: the
mutation rate $\mu$ depends on time as 
\begin{equation}
  \mu(t) = \frac{\mu_0 -\mu_{\infty}}{2}
  \left(1-\tanh\left(\frac{t-\tau}{\delta}\right)\right) +
  \mu_{\infty},
\end{equation}
which roughly corresponds to keeping $\mu=\mu_0$ (a high value, say
$10/N$) up to a time $\tau-\delta$, then decrease it  up to the
desired value $\mu_{\infty}$ in a time interval $2\delta$ and continue
with this value for the rest of simulation. 

The limiting case of this procedure is to use $\tau=\delta=1$ and
$\mu_0=1$, which is equivalent to starting from a random genetic
distribution. Simulations show that all initial conditions tend to the
same asymptotic one, and that the variance reaches its asymptotic value
very quickly.

The assortativity is introduced through the mating range $\Delta$ which
represents the maximal phenotypic distance still compatible with
reproduction.  The reproduction phase is thus performed in this way.
We choose one parent at random in the population, while the other
parent is chosen among those whose phenotypic distance from the first
parent is less than $\Delta$.  The genome of the offspring is built
by choosing for each \emph{locus} the allele of the first or second
parent with the same probability and then  mutations are introduced
by inverting the value of one bit with probability $\mu$. In our
model we therefore assume absence of \emph{linkage}, which is a
simplification often used in literature. It must be remembered,
however, that this simplification is only reasonable in the case of
very long genomes distributed on many  independent chromosomes. The effects of linkage disequilibrium will be considered in a future work.

In this work we are interested in studying the combined effects of competition and assortativity on speciation. The simplest and computationally most efficient choice is that of a \emph{frequency-dependent} but \emph{density independent} fitness function. This makes the evolution of population size decoupled from that of the frequency distribution of phenotypes as can be shown in the mean-field approximation~\cite{Bagnoli}. It is thus possible to study a fixed-size population.

In general, the number of individuals carrying phenotype $x$ at time
$t$  is denoted by $n(x,t)$, the total population size by
$N(t)=\sum_{x=0}^L n(x,t) $ (fixed in our simulations) and the 
distribution of phenotypes by $p(x,t)=n(x,t)/N$. The evolution
equation for the distribution $p(x,t)$  is:

\begin{equation}
\tilde{p}(x,t) = \sum_{y,z}p(y,t)p(z,t)W(x|y,z) 
\end{equation}

\begin{equation}
 p(x,t+1) = \frac{A(x,t)}{\bar{A}(t)}\tilde{p}(x,t) 
\end{equation}
where $\tilde{p}(x,t)$ is the frequency of phenotype $x$ after the recombination and mutation steps, $W(x|y,z)$ is the conditional probability that phenotype $x$ is produced by parents with phenotype $y$ and $z$, $A(x,t)$ is the survival factor of phenotype $x$ at time $t$ and  $\bar{A}(t) = \sum_{x}A(x,t)\tilde{p}(x,t)$.
The survival factor is thus the unnormalized probability of surviving to the reproductive age. 
The idea beyond this approach is quite simple: individuals
with a survival factor higher than average have  the best chances to survive.

In general, the survival factor $A(x,t)$ depends on time either because of environmental effects (say, daily oscillations, not considered here) or because the chances of surviving depend on the competition with the whole population, i.e. 
\begin{equation}
A(x,t) = A(x,\tilde{\mathbf{p}}(t)).
\end{equation} 
In the presence  of competition, a non-overlapping generation model
is much  faster to simulate, since in this case we do not need to
recompute the survival factor for each individual, but only once per
phenotype per generation.

Since $A$ is proportional to the survival probability, and the probability of independent events is the product of the probabilities of the single events, we define :
\begin{equation}
 A(x,\tilde{\mathbf{p}}(t)) = \exp (H(x,\tilde{\mathbf{p}}(t))) 
\end{equation}
and we call $H$ the \emph{fitness landscape}, or simply fitness. 
In this way the events contribute additively to $H$. This choice is a common one (see for instance Refs. ~\cite{FitnessLandscape,Peliti1,Peliti2}) but
someone may prefer $A$ as the real fitness. 
Notice that  $A$ is defined up to a multiplicative constant factor, and so $H$  is defined up to an additive constant (thus it may always be made positive, consistently with the usual biological literature).

The fitness $H$ can be built heuristically. 
First of all there is the viability $H_{0}(x)$ of phenotype $x$, not depending on the interactions with other individuals. $H_{0}(x)$ can be therefore defined the \emph{static component} of fitness describing the adaptation to abiotic factors such as climate or temperature whose dynamics is much slower than that of a biological population (consider for instance the alternation of ice ages and warmer period in the geological history of earth). The next terms of the fitness function account for the pair interactions, the three-body interactions, etc. The general expression of the fitness landscape is thus:

\begin{equation}
  H(x,p) = H_0(x) + \sum_y H_1(x|y) p(y) + \sum_{yz} H_2(x|yz) p(y) p(z) + \dots
\end{equation}

In the present work we consider only the effect of the first two terms.
The static component
of the fitness is defined as:

\[ H_0(x) = e^{- \frac{1}{\beta} \left ( \frac{x}{\Gamma}
	 \right )^{\beta} } \]

\begin{figure}[ht]
\begin{center}
\includegraphics[scale=0.7]{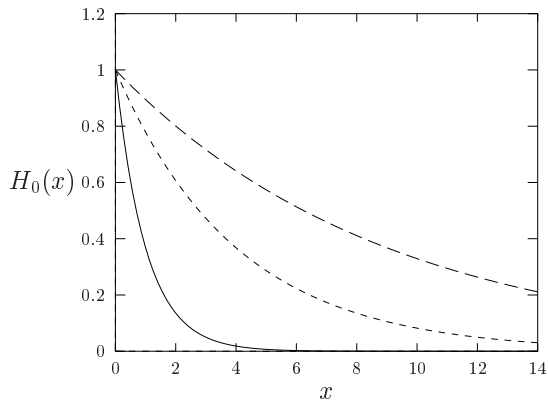}

\caption{Steep profiles of static fitness  $H_{0}(x)$. From top to bottom $\Gamma = 1,2,3$ and $\beta = 1 $ }

\label{fig:1}
\end{center}
\end{figure}

We choose this function because it can reproduce several landscapes
found in the literature by tuning the parameters $\beta$ and
$\Gamma$. In particular $H_0(x)$  becomes flatter and
flatter as the parameter $\beta$ is increased. When $\beta
\rightarrow 0$ we obtain the sharp peak landscape
 at $x=0$~\cite{Peliti1,Peliti2}; when
$\beta = 1$ the function is a declining exponential whose steepness
depends on the parameter $\Gamma$; and finally when $\beta
\rightarrow \infty$ the fitness landscape is constant in the range
$[0,\Gamma]$ and zero outside (step landscape). 
Some examples of the effects of $\Gamma$ on the static
fitness profile  are shown in figure~\ref{fig:1}.

The dynamic part of the fitness has a similar expression, with
parameters $\alpha$ and $R$ that control the steepness and range of
competition among phenotypes. 

The complete expression of the fitness landscape is:
 
 \begin{equation}
H(x,t) = H_0(x) - J\sum_y e^{- \frac{1}{\alpha} \left | \frac{x-y}{R} \right
|^{\alpha}}\tilde{p}(y,t)
\label{H(x)}
\end{equation} 
The parameter $J$ controls the intensity of competition with respect
to the arbitrary reference value $H_0(0)=1$.
If $ \alpha = 0$ an individual with phenotype $x$ is in competition only with
other organisms with the same phenotype; conversely
in the case $\alpha \rightarrow \infty$ a phenotype $x$ 
is in competition with all the other phenotypes in the range $[x - R,
x + R]$,  and the boundaries of this competition intervals 
blurry when $\alpha$ is decreased.

Let us introduce  some properties of the selection phase (we do not consider here the effects of recombination in the reproductive phase) by means of a simple examples.
Let us start with the case of a population in which all genotypes differ only for neutral genes (i.e. genes that do not affect $H$).
In this way, also after recombination the population is phenotypically homogeneous. In this case the survival factor $A(x, \tilde{\mathbf{p}}(t))$ is constant and equal to $\bar{A}$ and there is no selection at the distribution level. Indeed, the total number of individuals sharing the phenotype $x$ may be affected by competition, and the actual number of offsprings may be reduced so that the whole population is lead to extinction (which is not considered in our model, since we work at fixed population). But selection is not able to alter the phenotypic/genotypic distribution, since all phenotypes experience the same level of selection.

We now consider the case of a population composed of two phenotypes only, $x_1$ and $x_2$ with respective frequencies $\tilde{p}(x_1)=q$ and $\tilde{p}(x_2)=1-q$. Suppose that the two peaks are far enough in the phenotypic space that there is only intra-specific competition (already considered by the normalization of the probability distribution) and not interspecific competition; also assume that the static fitness landscape is flat so that $H_{0}(x) =1$ for both phenotypes. Under these conditions $A(x_1, \tilde{\mathbf{p}}(t)) = \exp (1- Jq)$;  $A(x_2,\tilde{\mathbf{p}}(t)) = \exp (1- J(1-q))$, and $\bar{A} = A(x_1, \tilde{\mathbf{p}}(t))\tilde{p}(x_1) + A(x_2,\tilde{\mathbf{p}}(t))\tilde{p}(x_2)=q\exp (1- Jq)+(1-q)\exp (1-J(1-q))$. The evolution equation \eqref{p} has only one free component
\begin{equation}
  p'(x_1) = q' =  \dfrac{A(x_1,\tilde{\mathbf{p}}(t))}{\bar{A}}\tilde{p}(x_1)=\dfrac{q\exp(-Jq)}{q\exp (- Jq)+(1-q)\exp (-J(1-q))} 
\label{p}  
\end{equation}
which exhibits fixed points for $p=0$, $p=1$ and $p=p^*=0.5$. For $J>0$ the only stable point is $p=p^*$, corresponding to the minimum competition felt by each strain. Notice that, differently from what intuition suggests, the  factor $F(q)=A(x_1,\tilde{\mathbf{p}}(t))/\bar{A}$ is not a monotonically decreasing function of $q$, but exhibits a minimum at $q\simeq 0.8$. Indeed, without mutations, $q=1$ (only one species) has to be a fixed point (no new species can be generated), so  $F(1)=1$, and in order to have a stable fixed point at $q=q^*=0.5$ ($F(0.5)=1$), one needs $F(q)<1$ for $0.5<q<1$. Similarly, one needs $F(q)>1$ for $0<q<0.5$.

We now briefly review the implementation of our model. The initial population is chosen at random and stored in a phenotype distribution matrix with \mbox{$L+1$} rows and $N$ columns. Each row represents one of the possible phenotypes; as the whole population might crowd on a single phenotype, $N$ memory locations must be allocated for each phenotype. Each generation begins with the reproduction step. The first parent is chosen at random; in a similar way, the second parent is randomly chosen within the mating range of the first one, \emph{i.e.} within the range $[\max \{ 0,x-\Delta \}, \min \{L,x+\Delta \}]$ where $x$ is the phenotype of the first parent.

The offspring is produced through uniform recombination, \emph{i.e.}, for each \emph{locus} it will receive the allele of the first or second parent with equal probability; the recombinant then undergoes mutation on a random allele with probability $\mu$. The newborn individuals are stored in a copy of the phenotype distribution matrix with $L+1$ rows and $N$ columns. At this stage we compute the survival factor $A(x,\tilde{\mathbf{p}}(t))$ for each phenotype and the average $\bar{A}$. The reproduction procedure is followed by the selection step. As we consider a constant size population, a cycle is iterated until $N$ individuals are copied back from the second to the first matrix. In each iteration of the cycle, an individual is chosen at random and its relative fitness is compared to a random number $r$ with uniform distribution between $0$ and $1$: if $r < A(x,\tilde{\mathbf{p}}(t))/\bar{A}(t)$ the individual survives and is passed on to the next generation, otherwise a new attempt is made.

It should be noted that both reproduction and selection steps are affected by stochastic components, \emph{i.e.} the reproduction and survival possibility of an individual does not depend only on its fitness but also on accidental and unpredictable circumstances, which is quite realistic. Consider for instance, an individual colonizing a new territory: its fitness will be very high due to the availability of resources and lack of competition, but, as the region is still scarcely populated, it may be difficult to find a partner and it may not reproduce at all. Similar remarks apply to an individual with high fitness that is accidentally killed by a landslip or by the flood of a river.

Another  interesting research problem that we address in this paper is that of the direction of evolution. It is common belief that evolution realizes a continuing advance towards more sophisticated forms. Recent advances in evolutionary theory (such as the theory of \emph{punctuated equilibrium}) and observations of evolutionary phenomena, however, seem to indicate that evolution is a  largely unpredictable, chaotic and contingent series of events, where small fluctuations may lead to major catastrophes that change the future course of development. Fisher's fundamental theorem of natural selection~\cite{Fisher1,Fisher2,Fisher3} and Holland's schemata theorem~\cite{Holland} seem to identify an evolutionary equivalent of free energy in the average fitness of the population. 

However,according to the modern interpretation of Fisher's theorem by Price~\cite{Price} and Ewens~\cite{Ewens} the total change in mean fitness is the sum of a partial change related to the variation in genotypic frequencies (operated by natural selection) and a partial change related to environmental deterioration, and Fisher's theorem predicts that the first contribution only is always non-negative. The schemata theorem on the other hand, states that schemas with higher than average fitness tend to receive an exponentially increasing number of samples in successive generations. As a consequence, the average fitness of the population is expected to increase as time goes by. Holland's theorem, however, is not general in that it is based on the assumption of fixed fitness for each phenotype. In realistic situations, fitness is frequency dependent, so that as the population becomes richer and richer in fitter individuals the intensity of competition also increases therefore counteracting a further increase in the frequency of the best adapted phenotypes. This increase in competition intensity is exactly what was meant by \emph{deterioration of the environmental conditions} in Price and Ewens's reformulation of Fisher's theorem~\cite{Price,Ewens}. In our formalism, the \emph{deterioration} is explicitly introduced by the competition term.

\section{Results.}

We have performed a large series of simulations in order to check the role of the various parameters. Since the detailed discussion of the parameter effects is rather long it is reported in the Appendix and we summarize here the main results. The numbering of the list corresponds to sections the the Appendix.

\def\theenumi{A.\arabic{enumi}}

\def\theenumii{\theenumi.\arabic{enumii}}
\def\theenumiii{\theenumii.\arabic{enumiii}}
\begin{enumerate}
  \item Flat static fitness landscape:
  \begin{enumerate}
	\renewcommand{\makelabel}[1]{\theenumii}
    \item Effects of assortativity:
    \begin{enumerate}
      \item Random mating: in the absence of assortativity speciation is not possible even if competition is very intense. All phenotypes are present although the distribution shows peaks and valleys.
      \item Moderate assortativity: the distribution becomes sharp, all but one phenotypes disappear even in the absence of competition.
      \item Strong assortativity (and absence of competition): coexistence of multiple species is possible, but this is not a stable distribution due to random fluctuations.
      \item Maximal assortativity: similar to previous case but transients with many coexistent phenotypes. 
    \end{enumerate}
    \item Effects of competition:
    \begin{enumerate}
      \item Weak competition in the presence of strong assortativity: stable coexistence. 
      \item Deterioration of the environment and Fisher's theorem: effective fitness decreases during simulations due to competition.
      \item Role of competition range: as the range of competition increases the number of coexisting species decreases. 
      \item High competition level and short range: increasing the competition intensity may increase the number of coexisting species. 
      \item Interplay between mating range and competition: an increase in competition may counteract an increase in the mating range. 
    \end{enumerate}
  \end{enumerate}
  \item Steep static fitness landscape: 
  \begin{enumerate}
 	\renewcommand{\makelabel}[1]{\theenumii}
   \item Absence of competition: only one species. 
    \item Stabilizing effects of competition: a moderate competition level may induce coexistence. 
    \item Competition-induced speciation: higher competition levels correspond to more species. 
    \item Interplay between mating range and competition: an increase in competition may counteract an increase in the mating range also in the case of steep fitness. 
  \end{enumerate}
  \item Influence of genome length: by repeating the simulations
  after doubling the genome length we show that, after rescaling the competition and assortativity range with the genome length, the final population distribution also scales linearly. 
\end{enumerate}

From these simulation emerges that the relevant parameters for the speciation (for a given static fitness landscape) are the competition intensity and assortativity (mating range). In the presence of competition, mutations play almost no role.

\section{Phase diagrams.}

One of the purposes of the present work was to study the evolutionary
dynamics in the widest possible range of competition and
assortativity so as to bring to light a possible synergetic effect on
speciation. To be consistent with the intuitive idea that higher assortativity corresponds to shorter mating ranges, we define the assortativity as: $\mathcal{A} = L - \Delta$. 
  As this kind of research requires a huge number of
simulations, the problem arises to find an easily computable
mathematical parameter  suitable for monitoring the speciation
process.

With this respect, one of the first candidates is the
\emph{variance} that, as is common knowledge, represents a measure of
the dispersion of a distribution:

\[var = \sum_i (x_i - \bar{x})^2p(x_i) \]

The simulations, in fact, show that, as
competition and/or assortativity is increased, the frequency
distribution first widens, then it becomes bimodal and eventually it
splits in two sharp peaks that move in the phenotypic space so as to
maximize their reciprocal distance: each of this steps involves an
increase in the variance of the frequency distribution.

It may be argued that the choice of variance as a parameter to monitor speciation may lead to interpretation mistakes. As an example, in a regime of random mating and high competition intensity, a trimodal distribution spanning the whole phenotypic space arises (when the static fitness landscape is flat): such distribution, shown in Figure~\ref{fig:2}, features a very high variance value. On the other hand, we also showed that in the absence of competition ($J = 0$), a speciation event may be induced by very high assortativity ($\Delta = 0$, $\Delta = 1$), but one of the newborn species soon disappears due to random fluctuations: the final distribution is therefore represented by a single delta-peak whose variance is obviously zero. It is therefore true that the variance does not allow to detect transient speciation events, but actually in the present work we are mainly concerned with the stable coexistence of the new species. In the interpretation of the variance values of the two examples we are discussing, therefore, we will not conclude that the former is closer to speciation than the latter but we will deduce that a trimodal distribution with three not completely split humps is closer to species coexistence than a single delta peak even if that represents the surviving species of a recent speciation event.

Before describing the plots of variance as a function of competition and assortativity, it is necessary to underscore that  for each couple of parameters $J,\mathcal{A}$ of the phase diagrams, the variance was averaged over 10 independent runs, so as to account for the stochastic factors influencing the evolutionary patterns while the plots shown in the preceding sections represent the output of a single run of our program.

As we will discuss in more detail, when variance is plotted as a
function of both competition and assortativity, the surface shows a
sharp transition from a very low value to a plateau at  a very high
variance level when assortativity and competition become larger than
a certain threshold. Analysis of the distribution shows that the sharp
transition from the low to the high variance level is indicative of
the shift from a state with a single quasi-species to a state with
two or more distinct quasi-species. The variance plot can be therefore
considered as a speciation phase diagram.

\begin{figure}[ht!]
\begin{center}

\includegraphics{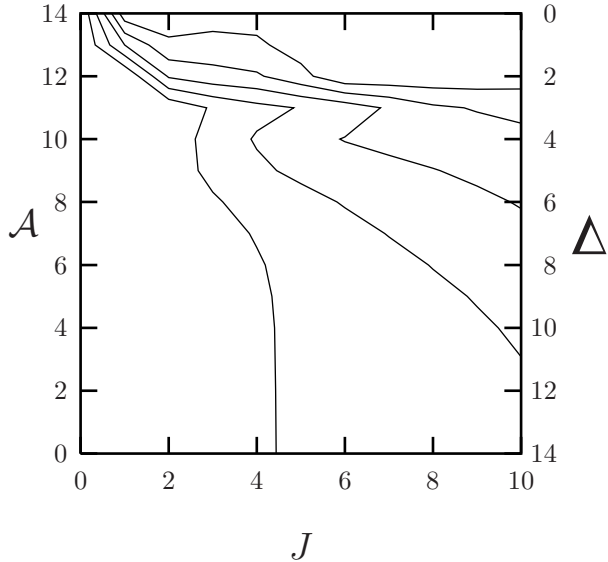}

\caption{Contour plots $var = 5,10,15,20,25$ for flat static fitness landscape. Parameters: $\beta = 100$, $\Gamma = 14$, $\alpha = 2$, $R = 2$, $p = 0.5$. Annealing parameters: $\mu_0 = 10^{-1}$, $\mu_{\infty} = 10^{-6}$, $\tau = 10000$, $\delta = 3000$. Total evolution time: 30000 generations. Each point of the plot is the average of 10 independent runs.  }
\label{phase-flat}
\end{center}
\end{figure}

Let's start by examining the variance plot as a function of
competition and assortativity in the relatively simple case of a flat
static fitness.

A graphical way to illustrate the synergistic effect is to study
the contour plots of the tridimensional speciation phase diagram
(Figure~\ref{phase-flat}).

The contour plots divide the $J$, $\mathcal{A}$ plane in three regions: the area on the left corresponds to the state with a single quasi-species, the area near the upper-right corner corresponds to the state with two or more distinct quasi-species.
The area on the right below the previous one corresponds to wide distributions, with peaks which are not completely isolated. The phase transition is located in correspondence of the
wiggling of contour lines for $\mathcal{A}\simeq 11$ ($\Delta \simeq 3$).

If a point, owing to a change in competition and/or assortativity, crosses these borderlines moving from the first to the second region, a speciation event does occur. It should be noted that in the case of flat static fitness, and, to a smaller extent, also in the case of steep static fitness, in the high competition region the contour plots tend to diverge from each other showing a gradual increase of the variance of the frequency distributions. This is due to the fact that, even if competition alone is not sufficient to induce speciation in recombinant populations, it spreads the frequency distribution that becomes wider and wider, and splits into two distinct species only for extremely high assortativity values. In this regime of high competition only the ends of the mating range of a phenotype $x$ are populated and the crossings between these comparatively different individuals will create once again the intermediate phenotypes preventing speciation until assortativity becomes almost maximal.  

It can be noted that in the absence of competition, however assortativity is increased, the variance of the final distribution will always be zero. In a random mating regime ($\mathcal{A}=0$), in fact, the initial distribution remains bell-shaped until $T \cong \tau$ and then, due to recombination it shrinks in a delta peak occupying a random position in the phenotypic space: speciation does not occur. Conversely, with high assortativity ($\mathcal{A} = 13$, $\mathcal{A} = 14$) the initial distribution immediately turns into two delta peaks at $x=0$ and $x=14$ linked by a continuum of very low frequencies on the intermediate phenotypes. After $T \cong \tau$ two scenarios are possible: one of the delta peaks in $x=0$ and $x=14$ disappears and the other one survives, or, alternatively, both the extreme peaks disappear and a new peak arises in the central region of the space: a transient speciation event has occurred.

With a weak competition level such as $J=2$, even in the case of random mating, the distribution always remains bell-shaped and never turns in a delta peak: a weak level of competition is therefore necessary to stabilize a bell-shaped distribution. The situation does not change until $\mathcal{A} = 12$: the distribution splits in two peaks each one covering two phenotypes. When assortativity is further increased to $\mathcal{A} = 13$ and $\mathcal{A} = 14$ three and five stable delta peaks appear respectively, evenly spaced in the phenotypic space so as to relieve competition. These speciation patterns explain the abrupt increase in variance for high assortativity.

When $J=6$ the higher competition intensity changes the patterns: for $\mathcal{A} = 0$ the distribution is always bell-shaped, but with intermediate assortativity such as $\mathcal{A} = 10$ and $\mathcal{A} = 11$ the distribution becomes bimodal and remains such during all the simulation. When $\mathcal{A} = 12$ the final distribution is composed by three and not two peaks as in the case $0<J<6$: two delta peaks appear at the opposite ends of the space and one peak spanning two phenotypes arises in the central region. The higher competition intensity affects also the case $\mathcal{A} = 13$: two delta-peaks appear at the extreme ends of the phenotypic space, and two peaks each one covering two phenotypes appear in the central region (recall that only three delta peaks appeared for $J=4$). Finally, for $\mathcal{A} = 14$ we have the usual patterns with five evenly spaced delta-peaks.

These patterns basically do not change for higher competition values, the only differences being rather quantitative than qualitative. For instance when $J=10$ the distribution is bell-shaped for all assortativity values from $\mathcal{A}=0$ to $\mathcal{A} =6$ but the distribution is wider and thus the variance is larger than the cases with weaker competition. For $\mathcal{A}$ in the range from 7 to 11 the distribution becomes bimodal: the stronger competition therefore allows the distribution to become bimodal with much lower assortativity levels than those required with $J=6$. Finally when $\mathcal{A}=12$ three peaks appear but all of them and not only the intermediate one cover several phenotypes. No significant difference from the cases $J<10$ can be detected when $\mathcal{A} = 13, 14$.

The speciation phase diagram has been studied also in the case of a steep static fitness landscape. The two diagrams are qualitatively similar, except that now the multiple-species (large variance) phase extends to $\mathcal{A} \simeq 8$ ($\Delta\simeq 6$). On the other hand, the intermediate region of wide distributions shrinks.

\begin{figure}[ht!]
\begin{center}

\includegraphics{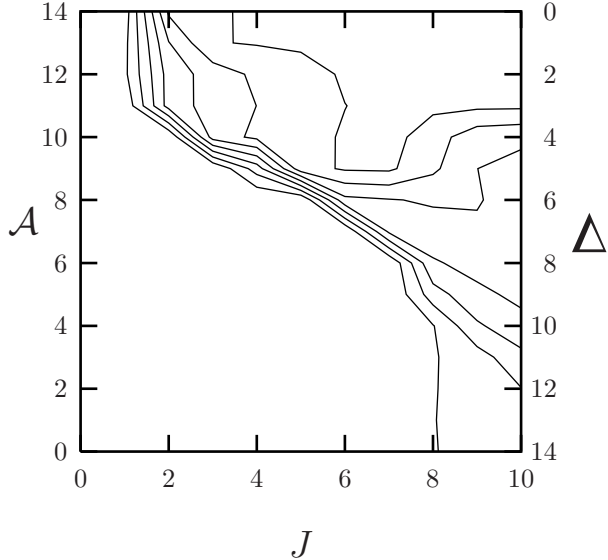}

\caption{Contour plots $var = 5,10,15,20,25,30,35$ for steep static fitness landscape. Parameters: $\beta = 1$, $\Gamma = 10$, $\alpha = 2$, $R = 4$, $p = 0.5$. Annealing parameters: $\mu_0 = 10^{-1}$, $\mu_{\infty} = 10^{-6}$, $\tau = 10000$, $\delta = 3000$. Total evolution time: 30000 generations. Each point of the plot is the average of 10 independent runs.  }
\label{phase-steep}
\end{center}
\end{figure}

As in the previous case, we analyze some
significant contour plots (Figure~\ref{phase-steep}). The down sloping
shape of these lines, again, is a strong indication of a synergistic
interaction of competition and assortativity on the speciation
process.
The contour plots show that for moderate competition there is a synergistic effect between competition and assortativity since a simultaneous increase of $J$ and $\mathcal{A}$ may allow the crossing of the borderline whereas the increase of a parameter at a time does not. On the other hand, for larger values of $J$ the phase diagram shows a reentrant character due to the extinction of one of the species that cannot move farther apart from the other and therefore cannot relieve competition anymore. It can also be noticed that for $J = 0$ the contour plot shows a change in slope due to extinction of one species owing to random fluctuations as shown earlier.

The following differences with respect to the case of flat static fitness can be detected: the curvature of the borderlines between the coexistence phases is higher, which indicates a stronger synergy between $\mathcal{A}$ and $J$; the absence of speciation for moderate $J$ here is not due to finite size effects. The contour plots of the phase diagram in the case of steep static fitness are shown in Figure~\ref{phase-steep}.

In order to make the interpretation of the phase diagram easier, we now briefly summarize the most significant evolutionary patterns for various values of $J$ and $\mathcal{A}$.

In a regime of random mating and absence of competition \mbox{($J=0$, $\mathcal{A}=0$)} the initial frequency distribution quickly turns in an asymmetrical bell-shaped curve in the neighborhood of $x=0$ which then becomes a delta-peak in $x=0$ after $T \cong \tau$. The situation is basically the same for any value of assortativity, the only partial exception being represented by the fact that for maximal assortativity ($\mathcal{A}=14$) the initial distribution immediately turns in a delta peak in $x=0$ followed by a tail of low frequency mutants. These mutants however disappear after $T = \tau$ so that the stationary distribution is still represented by a single delta-peak in $x=0$ and the variance equals zero.

We now describe what happens for $J=2$ so as to explain the abrupt increase in variance shown by the plot at high values of assortativity. For $\mathcal{A} < 11$ the initial distribution turns in a symmetrical bell-shaped distribution and then in a delta-peak in $x=0$ so that the variance will be equal to zero. When $\mathcal{A} = 11$ however, the initial distribution splits in two bell-shaped curves at the opposite ends of the phenotypic space that, after $T = \tau$ turn in two delta peaks in $x=0$ and $x=10$. This speciation event determines a sharp rise in variance from $var = 0$ to $var \cong 22$. A second jump in variance can be observed with maximal assortativity, $\mathcal{A} = 14$: in this case the initial distribution splits in two delta peaks in $x=0$ and $x=14$ linked by a continuum of low-frequency mutants; after $T = \tau$ the mutants disappear and three delta peaks appear at the opposite ends and in the middle of the phenotypic space. This is why the variance jumps to about 30.

When $J=4$ in the assortativity range 0-10 the initial distribution moves towards $x=0$ and then becomes a delta peak in this position. When $\mathcal{A} = 11$ the distribution splits in two bell-shaped curves one of which becomes a delta peak in $x=0$ and the other becomes a peak covering two phenotypes, usually $x=11$ and $x=12$. It should be noted that for $J=2$ the second species was represented by a delta-peak too; the higher competition experienced for $J=4$ however can be more easily relieved spreading the species over several phenotypes. The necessity to relieve the competition pressure also explains why for $\mathcal{A} = 13$ not two (as with $J=2$) but three delta-peaks appear in the final distribution. When $\mathcal{A} = 14$ however the final number of delta peaks is five as with $J=2$, probably because a higher number of peaks would make interspecific competition not sustainable with the competition range $R = 4$.

When $J=6$ the pattern is basically the same as with $J=4$ apart for a couple of differences: when $\mathcal{A} = 8$ the distribution instead of becoming a delta peak in $x=0$ splits in two bell-shaped distributions that after $T = \tau$ shrink and move towards the opposite ends of the space so that the variance can reach a final value of about 24; another important difference is represented by the fact that the appearance of a delta peak in $x=0$ and a peak covering $x=11,12,13$ now occurs with $\mathcal{A} = 9$ and not $\mathcal{A} = 11$ as with $J=4$ which represents a typical example of the synergetic interplay between competition and assortativity.

When $J=8$ in the range $\mathcal{A}=0$ to $\mathcal{A}=5$ the initial distribution does not become a delta peak in $x=0$ but it remains a bell-shaped (even if asymmetrical) curve with a variance around 4.5. As a consequence, the variance plot now increases more gradually and not in a stepwise manner as observed so far. When $\mathcal{A}$ is in the range 6-8 however, the variance jumps to 20-25 because the distribution splits in two bell-shaped curves as with $J=6$ could happen only with $\mathcal{A}=8$. When $\mathcal{A}=9$, contrary to what observed with $J=6$ no delta peak can appear in $x=0$, but the final distribution comprises a first peak covering phenotypes $x=0,1,2,3$ and a second peak spanning phenotypes $x=9,10,11,12,13$ so as to relieve the competition pressure. Finally, for $\mathcal{A} \ge 11$ three delta peaks appear at the ends and in the middle of the phenotypic space: this pattern for $0<J<8$ could be observed only with maximal assortativity $\mathcal{A} = 14$.

For $J=10$ the effects of this extremely high competition are most evident at low assortativity. When $\mathcal{A} = 0$ the initial distribution becomes wide and flat covering the phenotypic range 0-12 and after $T = \tau$ it becomes slightly bimodal. As a consequence, the variance is very high reaching $var \cong 10$.This pattern becomes more and more extreme as assortativity increases, until, for $\mathcal{A} = 4$ the distribution spans the whole phenotypic range 0-14 and it becomes more markedly bimodal with maximal frequencies on $x=0,1,2$ and $x=12,13$. The variance increases accordingly up to $var \cong 20$. For $\mathcal{A} >4$ the pattern is the same as for $J=8$.

\section{Conclusions.}

A microscopic model has been developed for the study of sympatric speciation \emph{i.e.} the origin of two reproductively isolated strains from a single original species in the absence of any geographical barrier.

In all our simulations we employed a simulated annealing technique, starting the run with a very high mutation rate that then is decreased according to a sigmoidal function which allows to attain the stationary distribution in a reasonably short runtime. 

We showed that in a flat static fitness landscape, assortativity alone is sufficient to induce speciation even in the absence of competition. This speciation event, however, is only transient, and soon one of the two new species goes extinct due to random fluctuations in a finite-size population. A stable coexistence between the new species, however, could be achieved by introducing competition. In fact, intraspecific competition turned out to stabilize the two groups by operating a sort of negative feedback on population size.

The simulations also showed that the assortativity level necessary for speciation could be reduced as competition is increased and \emph{vice versa} (except for the regime of extremely high competition), which strongly argues for a synergistic effect between the two parameters.

Similar patterns could be observed with a steep static fitness landscape. A high assortativity level is sufficient to induce speciation, but in the long run only the peak with maximal fitness  survives. The coexistence of the two species again is stabilized by competition.

A special attention was devoted to finite size effects. We showed that imposing maximal assortativity, in the presence of moderate competition, it was possible to reduce dispersion of offsprings and thus stabilize genotypically homogeneous peaks. In particular, in our 15-phenotypes fitness space, it was possible to stabilize the coexistence of three species, whose peaks tended to become symmetrical as the steepness of the fitness landscape was reduced.

We also showed that speciation has  the character of a phase transition, as the variance versus assortativity and competition surface shows a sharp transition from a low variance region corresponding to one species, to a high-variance region corresponding to two species. The curvature of the phase boundary once again supports the idea of a synergistic effect of competition and assortativity in inducing speciation. 
 
It is quite interesting to observe the behavior of the average of the
fitness landscape $\bar{H}$ with respect to time in the various
simulations. In general there is a large variation in correspondence
of the variation of the mutation rate. One can observe that there are
cases in which $\bar{H}$ increases smoothly when decreasing $\mu$,
while others, in correspondence of weak competition and moderate
assortativity exhibit a sudden decrease, often coupled to further
oscillations. At the microscopic level, this decrease corresponds to
extinction of small inter-populations that lowered the competition
levels.

We observe here a synergetic effect among mutation levels, finiteness
of population and assortativity. In an infinite asexual population,
assuming that there is a mutation level sufficient to populate each
phenotype, every variation of the population that would increase the
survival probability is always kept. Thus, an increase in the
mutation levels would lower $\bar{H}$, since the offspring
distribution is more dispersed of what would be the optimum. So, the
maximum of $\bar{H}$ is reached for $\mu \rightarrow 0$, with the
condition that there are still sufficient mutations to populate all
strains.

When population is finite this is no longer true. First of all there
are stochastic oscillations, and also the possibility of extinction
of isolated strains, which are hardly repopulated by the vanishing
mutations. Competition, by favoring dispersion, relieves this effect.

We observed that the decrease in mean fitness that sometimes can be
seen after the decrease of the mutation rate, is strongly affected by
assortativity in a non-linear way.  This lowering of fitness is due
to the  increase in intraspecific competition caused by  the
extinction of the low-frequency intermediate phenotypes. Maximal
assortativity in fact, prevents dispersion of offsprings so that
several peaks can appear in the middle of the phenotypic space
relieving intraspecific competition.  Slightly lower assortativity
values are too large to stabilize new peaks in the central regions of
the phenotypic space and too large to stabilize a wide distribution
so that they lead to a significant decrease in mean fitness.  A
further decrease of the assortativity, on the other hand, favors a
wide bell-shaped distribution so that the decrease in mean fitness
will be very modest.  

We have also checked that this scenario remains the same if all length quantities are rescaled (linearly) with the genome size. As a comparison, this finding does not hold when a non-recombinant population is in competition with a recombinant one~\cite{BagnoliGuardianiRicombinazione}, consistent with the fact that the genome of non-recombinant organisms is much shorter than that of recombinant ones. 

These patterns were observed treating the mutation rate as a tunable
parameter that changes during the simulation according to a sigmoid
function. It would be interesting to study the behavior of a
recombinant population when the mutation rate is an evolutionary
character. This will be the topic of a future work.

\section*{Acknowledgements.}
We acknowledge many fruitful discussions with the members of the CSDC and the DOCS group. 
We thank the anonymous referee for many interesting remarks and suggestions. 

\appendix

\section{Simulations.}

In order to check the dependence on the initial distribution, the frequency of phenotypes in the initial generation is binomially distributed:

\[ p(x) = \left ( \begin{array}{c}
                   L \\ 
		   x
		   \end{array} \right ) p^{x} (1 - p)^{L -x} \]

In the binomial distribution both the mean and the variance depend only on $L$ (the genome length) and $p$ (the probability of an allele being equal to 1):

\[ \bar{x} = Lp \]

\[ var = Lp(1-p) \]

For each parameter set we systematically performed simulations from five different initial distributions defined by $p = 0$, $p = 0.25$, $p = 0.5$, $p = 0.75$, $p = 1$. The situations $p = 0$ and $p = 1$ refer to the limit cases in which the population is completely concentrated on phenotypes $x = 0$ and $x = 1$ respectively, while in the $p = 0.5$ case the initial distribution is centered in the middle of the phenotypic space. Unless otherwise specified, the results of the simulations are independent from the initial distribution. This is valid also when different runs generate quantitatively different final distributions.

Another problem we addressed is that of finding the best  approximation of the stationary distribution of the population. Our model employs a finite population and is affected by several stochastic elements \emph{e.g.} in the choice of individuals in the reproduction and selection step, so that there will always be small fluctuations from generation to generation.
The presented results are obtained by averaging over a time interval able to average over these fluctuations but not hindering eventual drift effects.

\subsection{ Flat static fitness landscape.}

One of the simplest situations we can conceive, is a flat static
fitness profile in presence or absence of competition.

\subsubsection{Effects of assortativity.}

\paragraph{Random mating.}

\begin{figure}[ht]
\begin{center}
\begin{tabular}{cc}
\includegraphics[scale=0.6]{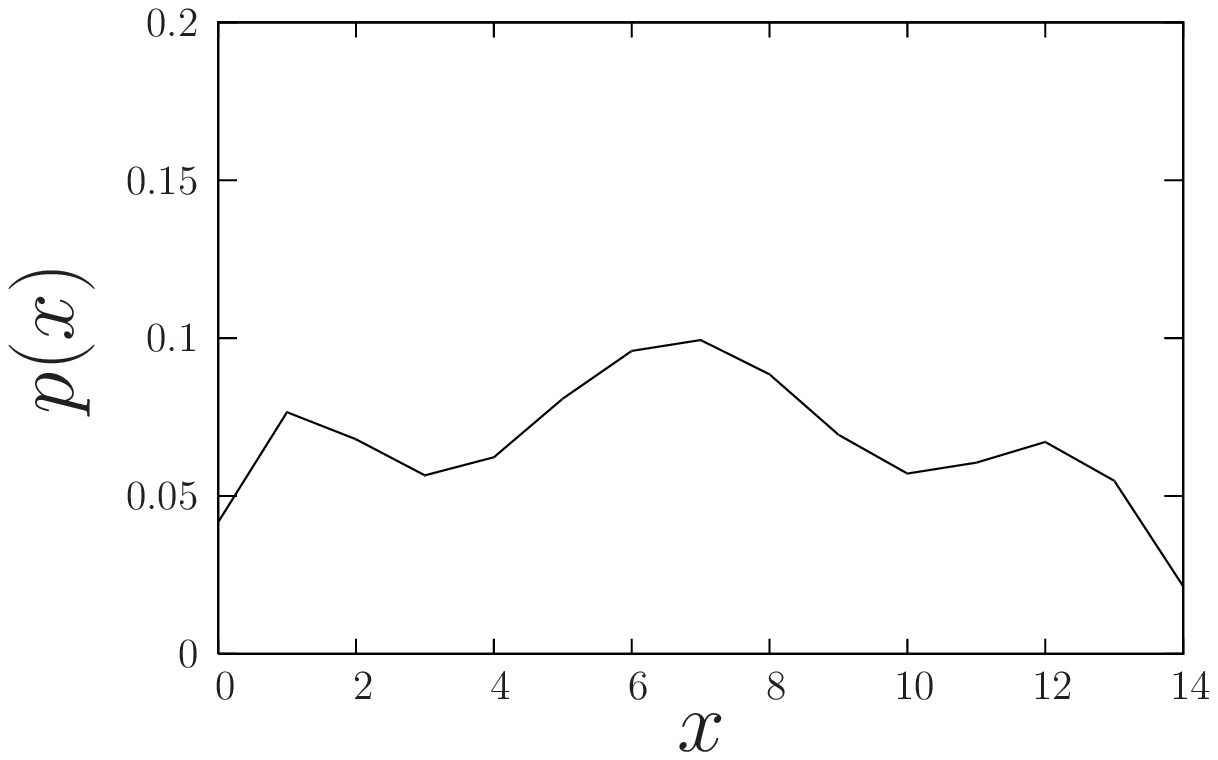} &
\includegraphics[scale=0.6]{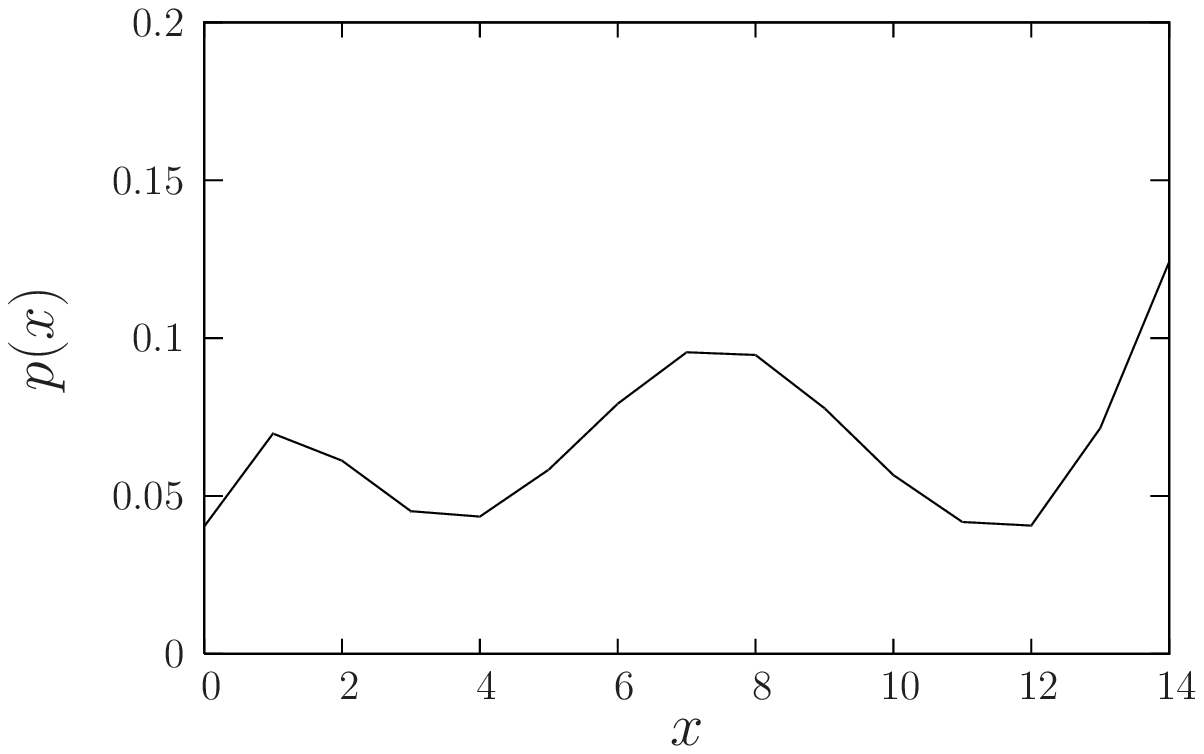}
\end{tabular}
\caption{A polymodal frequency distribution  generated in a random
mating regime ($\Delta = 14$) with  an extremely high competition intensity. Parameters: $J = 16$, $\alpha = 2$, $R = 7$. Left panel: trimodal distribution (averaged over generations 5000--5010); right panel: appearance of a transient peak in $x=14$ (averaged over generations 30000--30010). Except for the oscillations in the right tail, all other peaks and valleys are stable and localized.}

\label{fig:2}
\end{center}
\end{figure}

In this conditions, in a regime of \emph{random mating}, a population
is unable to speciate and, even employing
extremely high competition levels, the final state  is a trimodal frequency
distribution. The situation is shown in Figure~\ref{fig:2} (left panel).
Very early during the simulation, when the mutation rate is very high, two humps appear at the opposite ends of the phenotypic
space so as to minimize the mutual competition while the central hump
 is fed by the offsprings of crossings between the other
two humps. The continuous regeneration of the intermediate phenotypes as a result of the random mating, prevents the distribution from splitting into distinct species.   When $T \gg \tau$ the mutation rate becomes very low and transient peaks appear at $x = 0$ and $x = L = 14$ because these positions are very favorable in that individuals with these phenotypes experience a very low competition level. These peaks however are very short-lived and they appear and disappear very soon because of the dispersion of the offspring due to the random mating regime. In Figure~\ref{fig:2} we show the trimodal distribution that appears for $T \ll \tau$, and the distribution with a transient peak in $x=14$ appearing after $T = \tau$.

\begin{figure}[H]
\begin{center}
\begin{tabular}{cc}
\includegraphics[width=8cm,height=6cm]{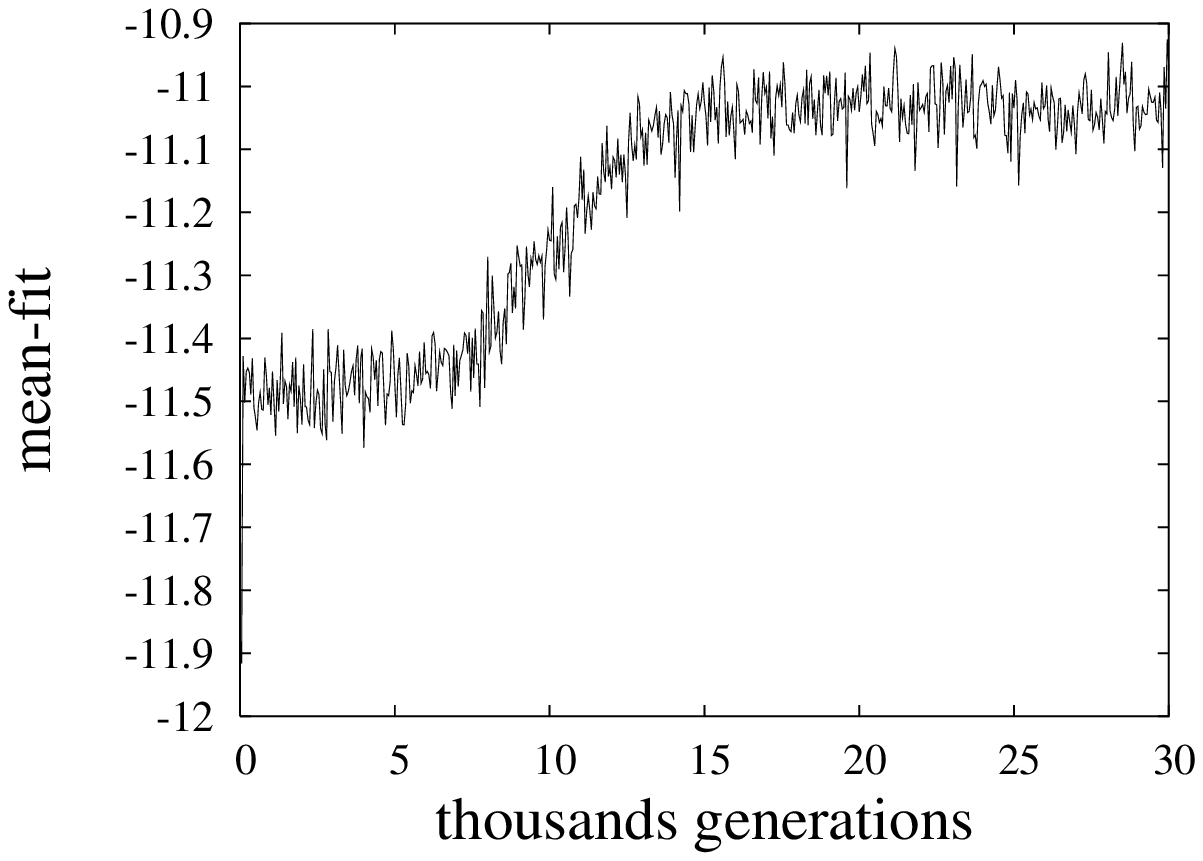} &
\includegraphics[width=8cm,height=6cm]{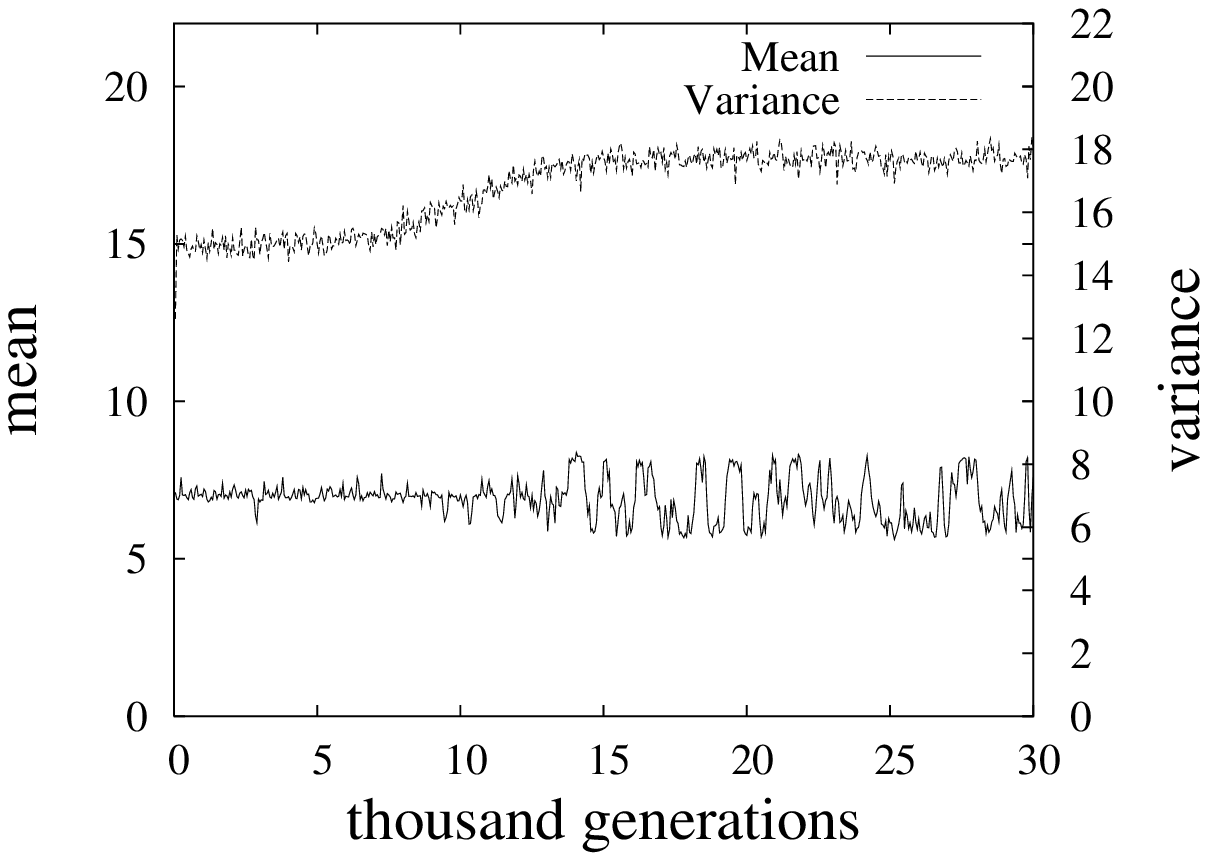}
\end{tabular}

\caption{ Average fitness, variance and mean in a regime of random mating ($\Delta = 14$) and strong competition ($J=16$, $\alpha = 2$,$R = 7$). Annealing parameters: $\mu_0 = 10^{-1}$, $\mu_{\infty} = 10^{-6}$, $\tau = 10000$, $\delta = 3000$. Total evolution time: 30000 generations. }

\label{fitness-variance_2}
\end{center}
\end{figure}

When $p = 0.5$ the initial distribution is centered at $x = L/2 = 7$ and it covers several phenotypes in the central region of the phenotypic space. The high mutation rate at the beginning of the simulation causes the distribution to become trimodal and extended to all the phenotypic space. This leads to an abrupt increase in the variance of the frequency distribution while the mean always oscillates around the same constant value $x = L/2 = 7$ because the deformation of the distribution is symmetric. The average fitness (Figure~\ref{fitness-variance_2}, left panel) also undergoes a significant increase because the population is now distributed on more phenotypes so that the number of individuals per phenotype is small and the competition level lowers accordingly. As already mentioned, for $T \gg \tau$ transient  peaks in $x=0$ and $x=L$ do appear, so that the mean of the distribution now shows wider oscillations  while the variance further increases because the formation of peaks in $x=0$ and $x=L$ implies the increase in frequency of phenotypes far from the mean of the distribution. The average fitness, on the other hand, also increases as a result of the increase of phenotypes $x=0$ and $x=L$ that experience very low competition. Figures~\ref{fitness-variance_2}  shows the plots of average fitness, variance and mean.

\paragraph{Moderate assortativity.}

\label{moderate-assort}
The scenario changes completely  if we impose a regime of
assortative mating. In this case, even in the absence of competition,
very interesting dynamical behaviors ensue.

As an illustration, let us consider the case $L = 14$.
If we set a moderate value of assortativity $\Delta=4$, the frequency
distribution progressively narrows until it becomes a sharp peak  at
the level of one of the intermediate phenotypes. This behavior can
be easily explained. Regardless of the value of the parameter $p$, due to the very high initial mutation rate,
the population becomes rapidly distributed following  a very wide and flat
bell-shaped frequency distribution whose average shows little oscillations in the phenotypic space.

The situation, however, changes significantly after $T = \tau$ when the mutation rate becomes very low. As an experimental observation, 
 the final frequency distribution is a delta peak typically located in one of the central phenotypes \footnote{Typically in the range 6 to 9 and less often also in $x=4$, $x=5$ and $x=10$}, the variety of positions being the result of the stochastic factors affecting the reproduction and selection steps.

\begin{figure}[H]
\begin{center}
\includegraphics[scale=0.6]{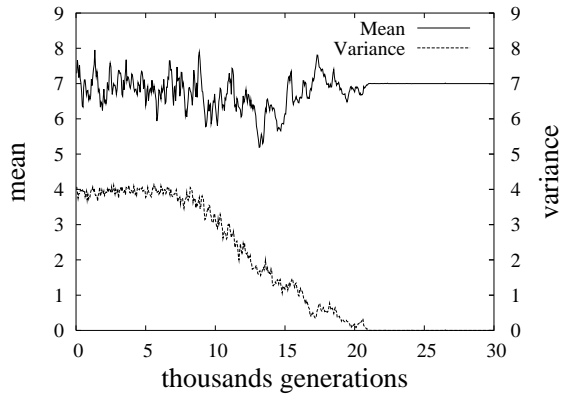}

\caption{Variance and mean in a regime of weak assortativity ($\Delta = 4$) and absence of competition ($J=0$). Annealing parameters: $\mu_0 = 10^{-1}$, $\mu_{\infty} = 10^{-6}$, $\tau = 10000$, $\delta = 3000$. Total evolution time: 30000 generations. }

\label{fitness-variance_3}
\end{center}
\end{figure}

Figure~\ref{fitness-variance_3} shows the plots of mean and variance  for this simulation.The final delta-peaks are also genotypically homogeneous (otherwise they will generate broader distributions in the next time step). Indeed, since the static fitness is flat, any genotypically homogeneous population is stable except for random sampling effects. 
  
 After the population has become genetically homogeneous, the only
fluctuations are  due to mutations, that sporadically generates
mutants. However, this effect cannot be seen in the Figure, due to
the scale of the axis and population size. The time eventually
required for a genotypic shift (fixation of a gene in the population)
is so long that this phenomenon could not be observed. 
In the presence of competition the gene fixation is extremely difficult due to the competition of the newborn mutants with the rest of population. 

The fact that this final delta-peak distribution is actually observed in simulations is due to the short mating range that tends to reproductively isolate  species. 
 
In the case of absence of competition the average fitness plot is not particularly interesting: the total fitness in fact, reduces to its static component which is equal to 1 for all phenotypes so that the average fitness plot will also be a constant. The variance plot, conversely, shows a decrease for $T > \tau$ which corresponds to the shrinking of the distribution from a bell-shaped curve to a delta-peak. 

As shown also in the phase diagram of figure~\ref{phase-flat},  there is a qualitative change 
of variance for $3\lesssim \Delta\lesssim 4$, for many values of $J$, indicating speciation. 
However,  in the absence of competition, this speciation effect is only transient, and the final distribution is a delta-peak. 

The mean plot, on the other hand, oscillates around a constant value with oscillations that become wider and more irregular during the transition from the bell-shaped to the delta-peak distribution.

\paragraph{Strong assortativity.}

\begin{figure}[ht]
 \begin{center}
\includegraphics[scale=0.7]{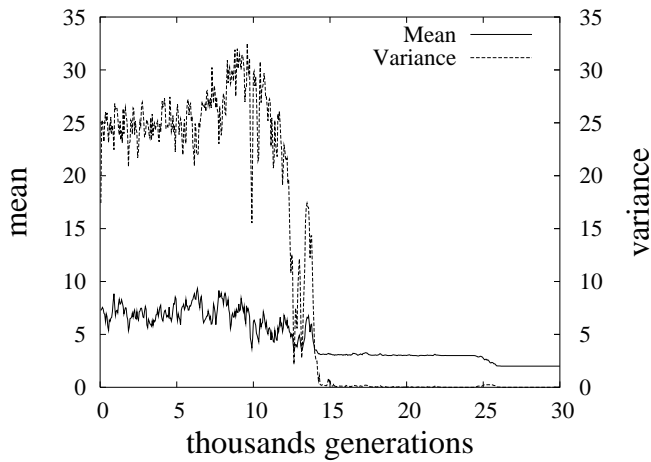}
  
\caption{Variance and mean in a regime of strong assortativity ($\Delta = 1$) and absence of competition ($J=0$). Annealing parameters: $\mu_0 = 10^{-1}$, $\mu_{\infty} = 10^{-6}$, $\tau = 10000$, $\delta = 3000$. Total evolution time: 30000 generations. The modest decrease in the mean that can be seen near the end of the simulation corresponds to a shift of the delta-peak from $x=3$ to $x=2$. }

\label{fitness-variance_4}
\end{center}   
\end{figure}

The results of a typical simulation with  a higher level of assortativity $\Delta=1$ are shown in Figure~\ref{fitness-variance_4}. At the beginning of the run, instead of a wide bell-shaped distribution (as in the case $\Delta = 4$), we find a distribution covering the whole phenotypic space with very low frequencies on the intermediate phenotypes and high frequencies on the extreme phenotypes in the regions around $x=0$ and $x=14$. This corresponds to an abrupt increase in variance, while the mean oscillates around the value $\bar{x} = 7$.

This particular distribution appears because the short mating range enforces matings only among very similar phenotypes. However, while matings among intermediate phenotypes with almost equal numbers of bits $1$ and $0$ produce offsprings different from the parents that are distributed over several phenotypes (thus lowering  the central part of the distribution), the matings between extreme phenotypes with a prevalence of zeros or ones do not scatter the offsprings that remain similar to the parents.
 When the mutation rate decreases for 
$T > \tau$, the scarcely populated intermediate phenotypes are removed by random fluctuations and only two delta peaks survive at the opposite ends of the phenotypic space (which is marked by another more modest increase in the variance around $T = \tau = 10000$): a speciation event has taken place. The coexistence of the two species however is not stable and later on one of the two peaks disappears still due to random fluctuations. As a result, the variance decreases abruptly to zero while the mean becomes constant. The final delta peak representing the stationary distribution is the survivor of the two peaks that had appeared at the opposite ends of the phenotypic space and for the same recombination effect discussed above, this asymmetric configuration \footnote{The positions of the final delta peaks are clustered in the intervals 2-4 and 11-12} persists forever.

\vspace{1 cm}

\paragraph{Maximal assortativity.}

\begin{figure}[ht] 
\begin{center}  Simulations
\includegraphics[scale=0.7]{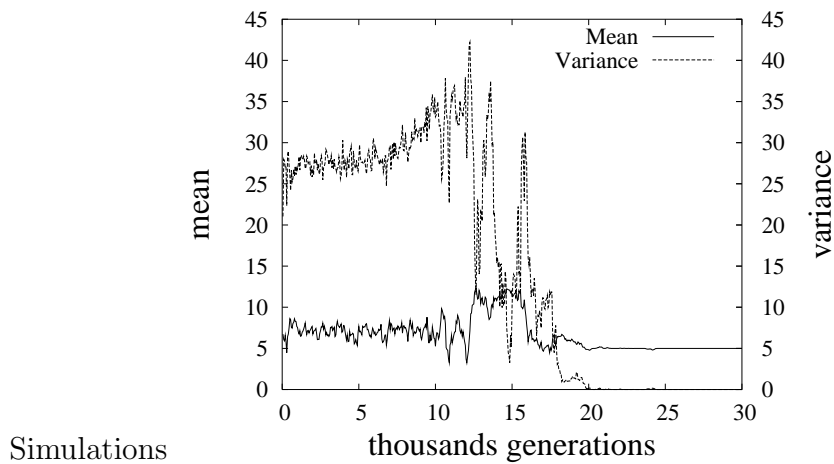}
     
\caption{Variance and mean in a regime of maximal assortativity ($\Delta = 0$) and absence of competition ($J=0$). Annealing parameters: $\mu_0 = 10^{-1}$, $\mu_{\infty} = 10^{-6}$, $\tau = 10000$, $\delta = 3000$. Total evolution time: 30000 generations. }

\label{fitness-variance_4bis}
\end{center}
\end{figure}

We finally explore the situation of maximal assortativity ($\Delta = 0$). For $T < \tau$ the situation is the same as with $\Delta = 1$: a distribution covering all the phenotypes, with low intermediate and high extreme values. For $T > \tau$ however, a sort of "bubbling" activity can be seen on the intermediate phenotypes with appearance and disappearance of many peaks. This is a consequence of the finite size of the population ($N = 3000$) that enables the scarcely populated intermediate phenotypes to be genotypically homogeneous; if we also consider that each individual can only mate with other individuals with the same phenotype it can be seen that there is no dispersion of offsprings so that the peaks of intermediate phenotypes will be hard to eradicate with random fluctuations. On the long run, however, only two peaks at the opposite ends of the phenotypic space will survive because their frequencies were high already at $T = \tau$. Finally one of the two peaks goes extinct and the stationary distribution is represented by a delta peak in one of the extreme regions of the phenotypic space. Figure~\ref{fitness-variance_4bis} shows the plots of variance and mean in a run with $\Delta = 0$:
it can be seen that the oscillations of both variance and mean for $T > \tau$ are wider and more irregular than in the case of Figure~\ref{fitness-variance_4} where $\Delta = 1$.

\vspace{1 cm}

\subsubsection{Effects of competition.}

The simulations just discussed show that assortativity alone is sufficient to induce speciation. The two newborn species, however, are not stable and soon one of them disappears due to random fluctuations.
The following experiments show that competition may stabilize multispecies coexistence.
With this respect, it must be remembered that competition is
inversely correlated with the phenotypic distance and therefore the
competition among individuals with the same phenotype is maximal.
Besides, competition is also proportional to the population density.
As a result, if the number of individuals of a species increases
(owing to random sampling, for instance), the intraspecific
competition increases as well, leading to a decrease in fitness
which, in turn, determines a reduction of the population size at the
following generation. In conclusion, competition acts as a stabilizing force preventing the
population from extinction.

\paragraph{Weak competition.}

We begin our discussion with the case of weak competition ($J = 1$, $\alpha = 2$, $R = 2$) and strong assortativity ($\Delta = 1$). At the beginning of the simulated annealing simulation, the distribution extends over all the phenotypic space with low frequencies on the intermediate phenotypes and high frequencies on the extreme ones because the latter prevalently produce offsprings similar to the parents while the former spread their offsprings over several different phenotypes. When $T = \tau$ the mutation rate significantly decreases and two different scenarios may arise:

\def\theenumi{\arabic{enumi}}

\begin{enumerate}

\item Stable coexistence of three species represented by delta-peaks located in the phenotypic space so as to maximize the reciprocal distance and thus minimize competition \footnote{The first one is typically in $x=0$, $x=1$, or $x=2$, the second one is in $x=6$, $x=7$, or $x=8$ and the third one is in $x=11$, $x=12$ or $x=13$.}.

\item Stable coexistence of two species located close to each other in the phenotypic space (5-6 phenotype units apart). This small distance is possible because competition is rather weak ($J = 1$, $\alpha = 2$, $R = 2$) \footnote{The first peak is typically in $x=3$ or $x=4$ while the second one is in $x=9$, $x=10$ or $x=11$.}.

\end{enumerate}

\begin{figure}[H]
\begin{center}
\includegraphics[scale=0.7]{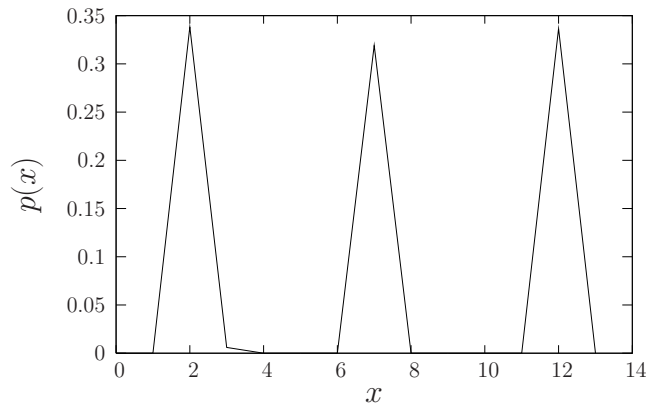}

\caption{The final distribution (generation 40000) obtained with weak  competition ($J = 1$, $\alpha = 2$, $R = 2$) and strong assortativity ($\Delta = 1$).}
\label{fig:4A}
\end{center}
\end{figure}

Figure~\ref{fig:4A} shows the final distribution (generation 40000) obtained in a run with $p=0$.

\vspace{1 cm}

The general idea is that competition stabilizes the coexistence and
prevent extinction due to random fluctuations. The mutation plays
almost no role in the presence of competition, except in offering
individuals the opportunity of populating empty "niches".  The
distribution reported in figure is stable because the peaks are
separated by a distance greater than the competition range.  If for
some reason, the central peak disappears and if one waits long
enough, then mutations  may repopulate this position (which is free
from competition), or  the other peaks may shift and occupy
intermediate positions, which is the  other stable configuration.
However, extinction of the central peak would be preceded by  a
decrease in its population, with a consequently decrease of
intraspecific competition,  and an increase the population of the
other peaks since we are working at fixed population. This in turn
would increase the intraspecific competition of the two external
peaks, thus lowering their fitness and this feedback  brings the
system back to the initial configuration.

\begin{figure}[ht]
\begin{center}
\begin{tabular}{cc}
\includegraphics[scale=0.6]{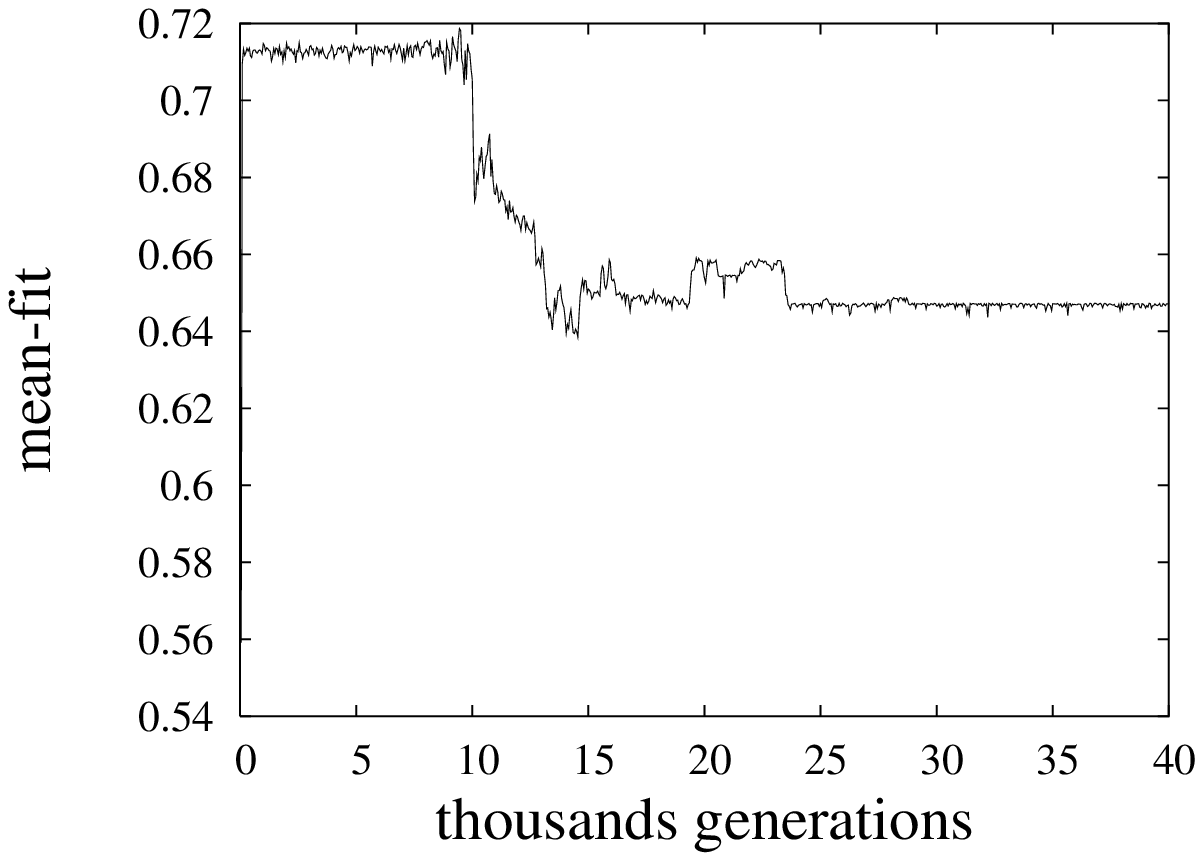} &
\includegraphics[scale=0.6]{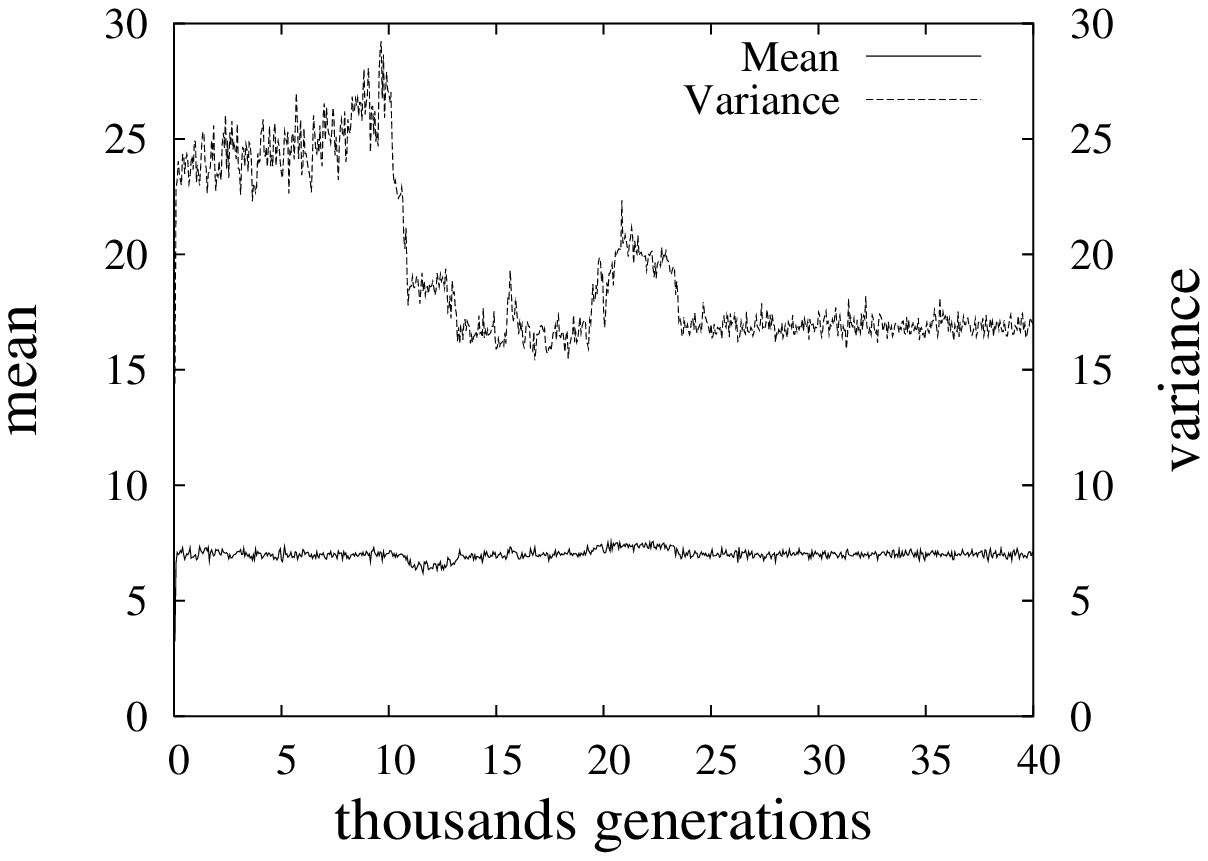}
\end{tabular}

\caption{ Average fitness, variance and mean in a regime of strong assortativity ($\Delta = 1$) and weak competition ($J=1$, $\alpha = 2$, $R = 2$). Annealing parameters: $\mu_0 = 10^{-1}$, $\mu_{\infty} = 10^{-6}$, $\tau = 10000$, $\delta = 3000$. Total evolution time: 40000 generations. }

\label{fitness-variance_4A}
\end{center}
\end{figure}

We now discuss the plots of average fitness, variance and mean. As $p=0$, the initial distribution is a delta-peak in $x=0$ that is very quickly deformed in a distribution spanning the whole phenotypic space; this obviously leads to an abrupt increase of the mean and variance of the distribution, but also the average fitness increases significantly because the spreading of the population over several phenotypes relieves competition. This effect obviously is stronger when $p=0$ and $p=1$ and this is why we are discussing the $p=0$ case. As shown in Figure~\ref{fitness-variance_4A}, the mean fitness oscillates at a very high level until $T = \tau$. After that, due to the decrease of the mutation rate, the individuals with intermediate phenotypes are removed by random fluctuations and the whole population will be concentrated on two delta peaks near the ends of the space. As a consequence of the very high intraspecific competition intensity in the two peaks, the average fitness drops abruptly. In this configuration the central part of the phenotypic space is not populated at all so that this is a particularly good location for a new colony to grow (very low competition). This is why two patterns may arise: the two peaks at the two ends of the phenotypic space may move closer to each other, or a third peak may appear in the middle of the phenotypic space. In the simulation we are describing these events are very fast and they do not exclude each other: the two peaks may move a little closer and then a third peak appears. The mean fitness however, does not change significantly: if a third peak appears, there is a decrease in the intraspecific competition that is partly balanced by an increase in the interspecific competition. The variance on the other hand will also decrease because the frequency of the central peak (that gives a very little contribution to the variance being near the mean) is higher than the sum of the old frequencies of the intermediate phenotypes. Finally, the mean of the distribution, jumps abruptly from $x=0$ to $x=7$ at the beginning of the run but then remains constant because all the following deformations of the distribution are roughly symmetrical.

\begin{figure}
\begin{center}
\centerline{\includegraphics{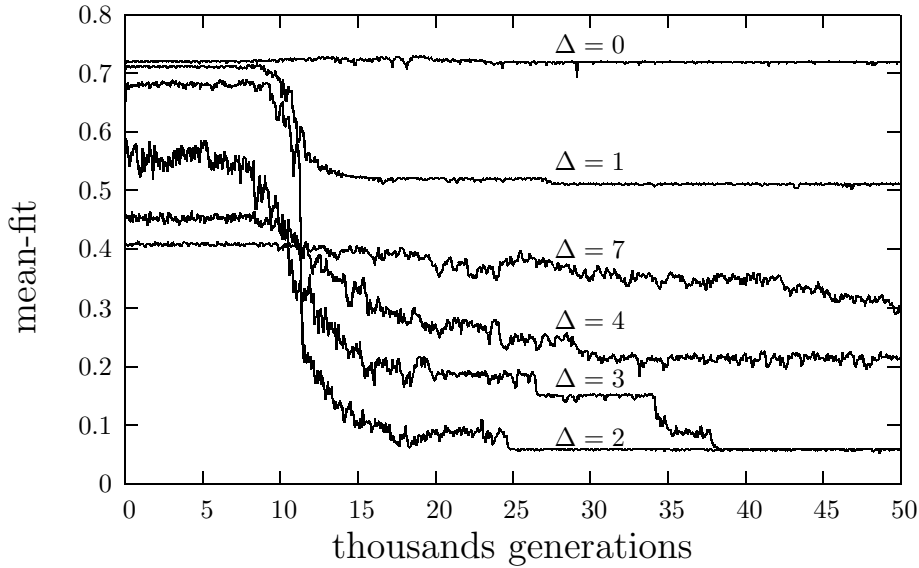}}
\end{center}
\caption{Average fitness plots for several values of assortativity $\Delta = 0,1,2,3,4,7$. Regime of weak competition: $J=1$, $\alpha = 2$, $R = 2$. Annealing parameters: $\mu_0 = 10^{-1}$, $\mu_{\infty} = 10^{-6}$, $\tau = 10000$, $\delta = 3000$. Total evolution time: 50000 generations. }
\label{fitness-plots}
\end{figure}

\paragraph{Deterioration of the Environment and Fisher's theorem.}
 As the decrease in mean fitness contradicts the traditional interpretation of Fisher's theorem, we further investigate this problem studying the role of assortativity. If assortativity is maximal ($\Delta = 0$) the mean fitness remains approximately constant throughout the simulation. After the extinction of the intermediate phenotypes  after $T \cong \tau$, we get a final distribution with four evenly spaced delta-peaks. This is possible because  $\Delta = 0$ forces individuals to mate only with partners showing the same phenotype. As we have observed earlier, genetic drift in finite populations tends to make them genetically homogeneous thus preventing dispersion of offsprings.  This is no longer true  when $\Delta = 1$ and no more than three delta-peaks can appear so that the intraspecific competition is high  and the mean fitness drops abruptly. The decrease in fitness is even more significant when $\Delta = 2$: the mating range is large enough to prevent the splitting of the initial distribution and we finally get a single peak spanning two phenotypes so that intraspecific competition is very high and fitness is minimal. The situation is slightly different for $\Delta = 3$: even if the final distribution is the same as for $\delta = 2$, this condition is reached about 15000 generations later because the larger mating range tends to keep a wide bell-shaped distribution. this is exactly what happens when $\Delta= 4$, when the large mating range stabilizes a wide bell-shaped distribution covering five phenotypes. The pattern becomes even more evident for $\Delta \geq 7$ when the final bell-shaped distribution spans seven phenotypes. The plots clearly show that as the bell-shaped distribution becomes wider and wider the decrease in fitness becomes less and less significant as a consequence of the reduced intra-specific competition. Figure~\ref{fitness-plots} shows the plots of mean fitness we have just discussed.

\paragraph{Role of competition range}

 In the next simulation we discuss the role of the competition range $R$, in particular, we consider the case $J = 1$, $\alpha = 2$, $R = 6$ with strong assortativity $\Delta = 1$ and a flat static fitness landscape $\beta = 100$, $\Gamma = 14$. We have seen that when the competition range is small (for instance $R = 2$ as in the experiment shown in Figures~\ref{fig:4A} and~\ref{fitness-variance_4A}, the stationary state is characterized by the coexistence of two or three species  close to each other in the phenotypic space. On the contrary, when $R = 4$ only two species can coexist and they are far from each other being located at the opposite ends of the phenotypic space \footnote{Typically, the first peak is found at $x=0$, and less often at $x=1$ while the second peak is usually at $x=14$ and less often at $x=13$.}. In Figure~\ref{fig:4A} we show the final distribution of a run with $p=0$.

\begin{figure}[ht]
\begin{center}
\includegraphics[scale=0.7]{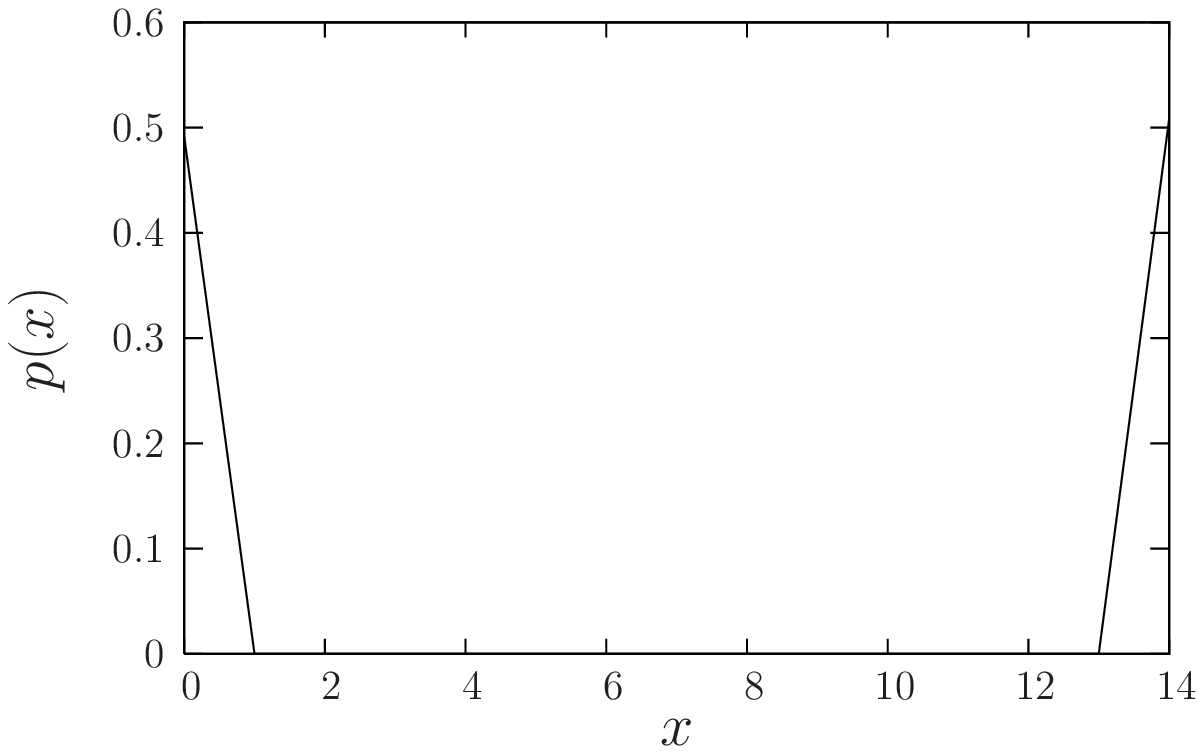}

\caption{The final distribution (generation 40000) obtained with weak  competition intensity but high competition range ($J = 1$, $\alpha = 2$, $R = 6$) and strong assortativity ($\Delta = 1$).}
\label{fig:4B}
\end{center}
\end{figure}

\begin{figure}[ht]
\begin{center}
\begin{tabular}{cc}
\includegraphics[scale=0.6]{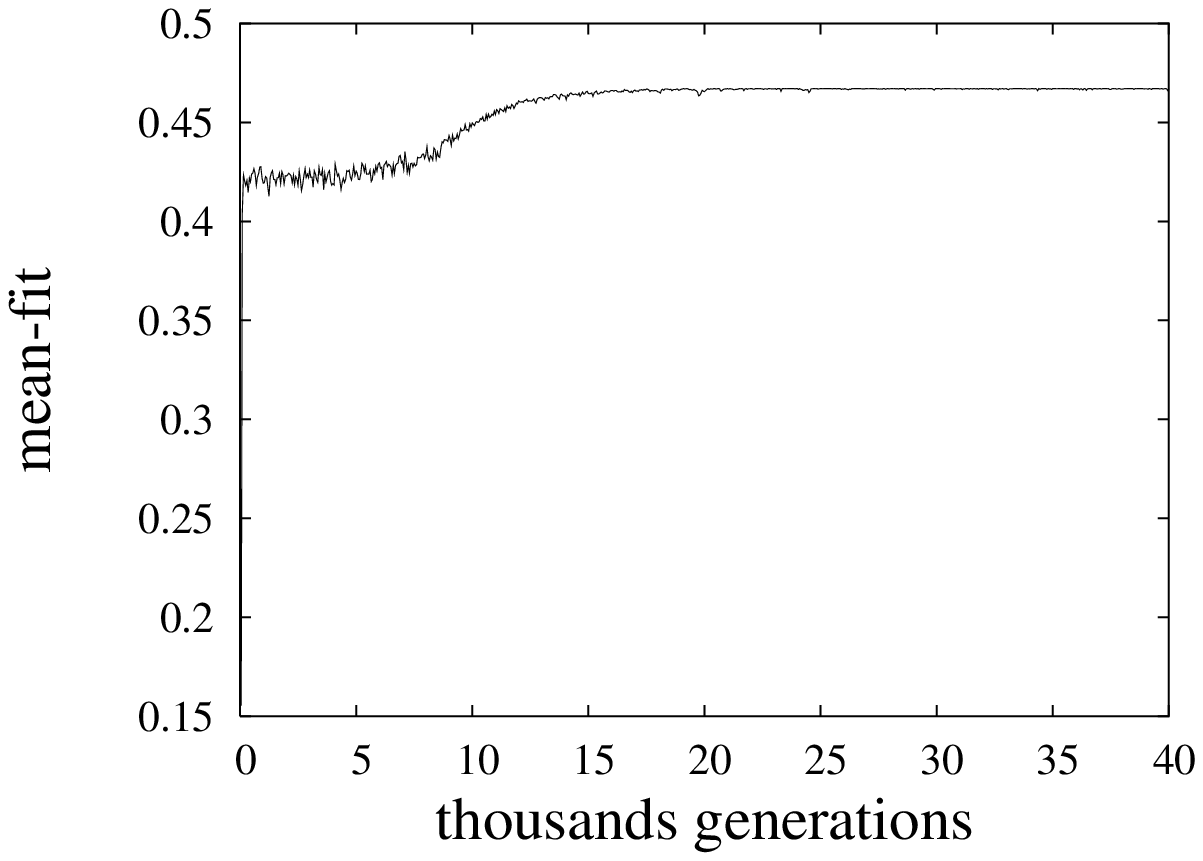} &
\includegraphics[scale=0.6]{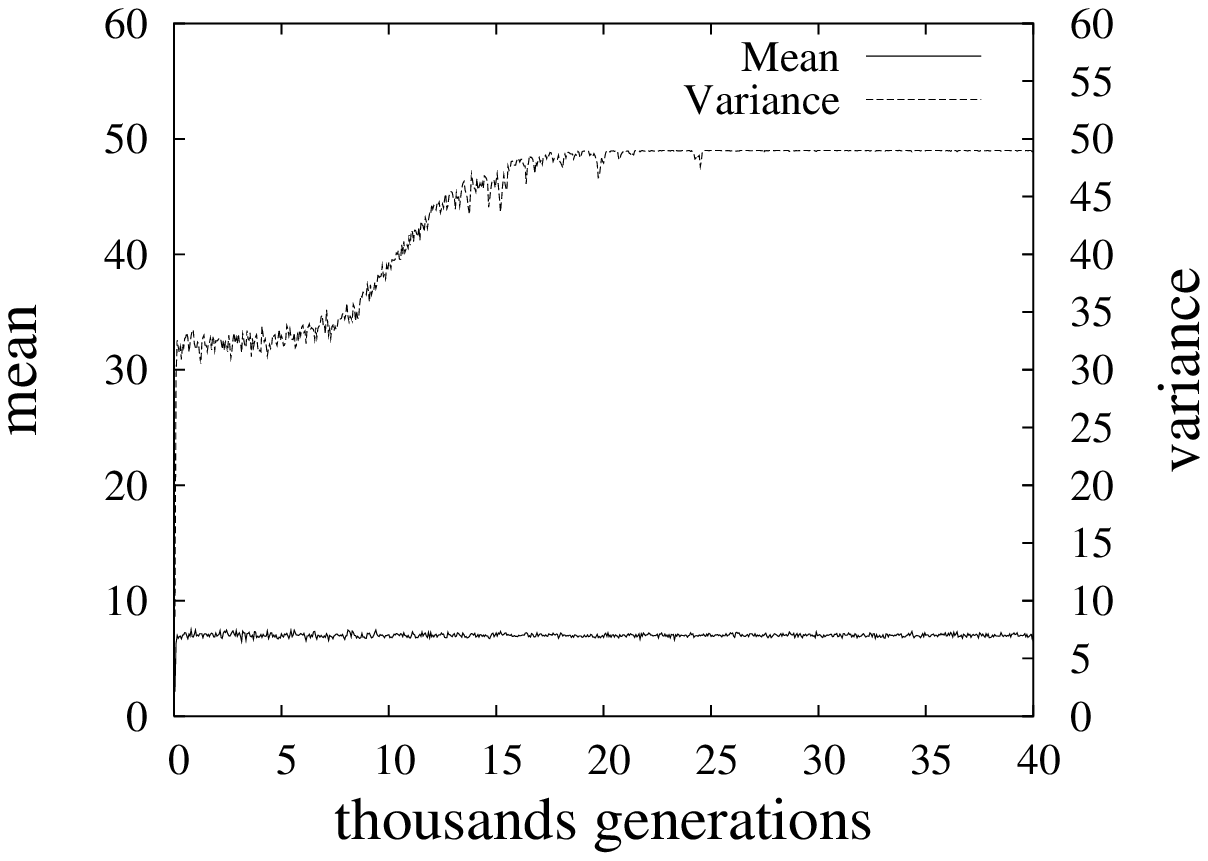}
\end{tabular}

\caption{ Average fitness, variance and mean in a regime of strong assortativity ($\Delta = 1$) and weak competition intensity but long competition range ($J=1$, $\alpha = 2$, $R = 6$). Annealing parameters: $\mu_0 = 10^{-1}$, $\mu_{\infty} = 10^{-6}$, $\tau = 10000$, $\delta = 3000$. Total evolution time: 40000 generations. }

\label{fitness-variance_4B}
\end{center}
\end{figure}

In this simulation the average fitness first increases abruptly when the initial delta peak in $x=0$ becomes a wide distribution covering all the phenotypic space then it oscillates around a constant value until $T = \tau$ and finally it increases again because when the whole population crowds in the two delta peaks at the opposite ends of the phenotypic space, the increase in intraspecific competition is outweighed by the fact that the distribution moves away from the central region of the space where the fitness is very low due to the high value of the competition range $R$. The variance also increases at the beginning of the run when the delta peak turns in a wide distribution, and it increases again when this distribution splits in two delta peaks far from the mean. Finally, the mean increases abruptly from $x=0$ to $x=7$ and then it oscillates around this value. Figure~\ref{fitness-variance_4B} shows the plots of mean fitness, variance and mean in a typical run.

\begin{figure}[ht]
\begin{center}
\includegraphics[scale=0.7]{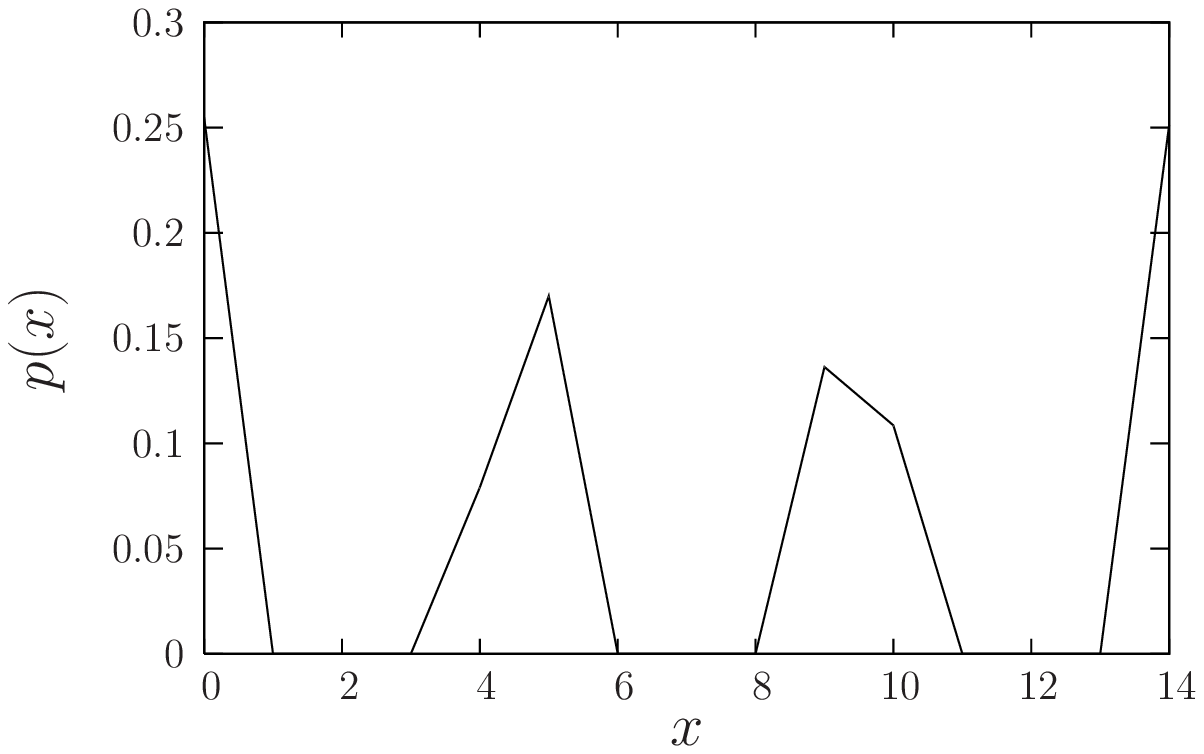}

\caption{The final distribution (generation 40000) obtained with strong  competition intensity but short competition range ($J = 8$, $\alpha = 2$, $R = 2$) and strong assortativity ($\Delta = 1$).}
\label{fig:4C}
\end{center}
\end{figure}

\paragraph{High competition.}
   We now consider the case of high competition intensity and small competition range. In particular, we will discuss a simulation with $J = 8$, $\alpha = 2$, $R =2$; for the sake of comparison with the previous simulations we will still keep $\Delta = 1$. When the competition was weak (see for instance Figure~\ref{fig:4A}) no more than three species could coexist in the phenotypic space. When $J=8$, however, the competition pressure is so strong that the population cannot be distributed only in three species and a fourth species appears so as to relieve competition. The simulations show that the first species is always in $x=0$ and the fourth is always in $x=14$. This is no surprise because these positions enjoy the lowest competition level; in fact the $x=0$ phenotype has got no competitors for $x<0$ and $x=14$ has no competitors for $x>14$. The second species sometimes appears in $x=5$ but more often it is represented by a peak covering phenotypes $x=4$ and $x=5$. In a similar fashion the third species sometimes appears in $x=9$, but typically it is represented by a peak spanning over phenotypes $x=9$ and $x=10$. The fact that a species comprises more than one phenotype again, is an evolutionary solution to relieve a competition that would be unbearable if the species was concentrated on a single phenotype. Figure~\ref{fig:4C} shows the final distribution of a run with $p=0$.

If we now analyze the mean fitness plot, we will notice that $H(x)$ is always negative whereas in Figure~\ref{fitness-variance_4A} it was always positive: the high intensity of competition thus represents a measure of the deterioration of the environmental conditions - to use the language of Price and Ewens reformulation of Fisher's theorem - that keep the fitness to a very low level. It will be noticed that the mean fitness increases rapidly at the beginning of the run because spreading the population over several phenotypes relieves the competition. The mean fitness then oscillates around a constant value and it grows again when the mutation rate decreases and the frequency distribution splits into four peaks. In this case, contrary to what shown in Figure~\ref{fitness-variance_4A} the decrease in intraspecific competition outweighs the increase in interspecific competition so that the fitness can increase.

\begin{figure}[ht]
\begin{center} 
\begin{tabular}{cc}
\includegraphics[scale=0.6]{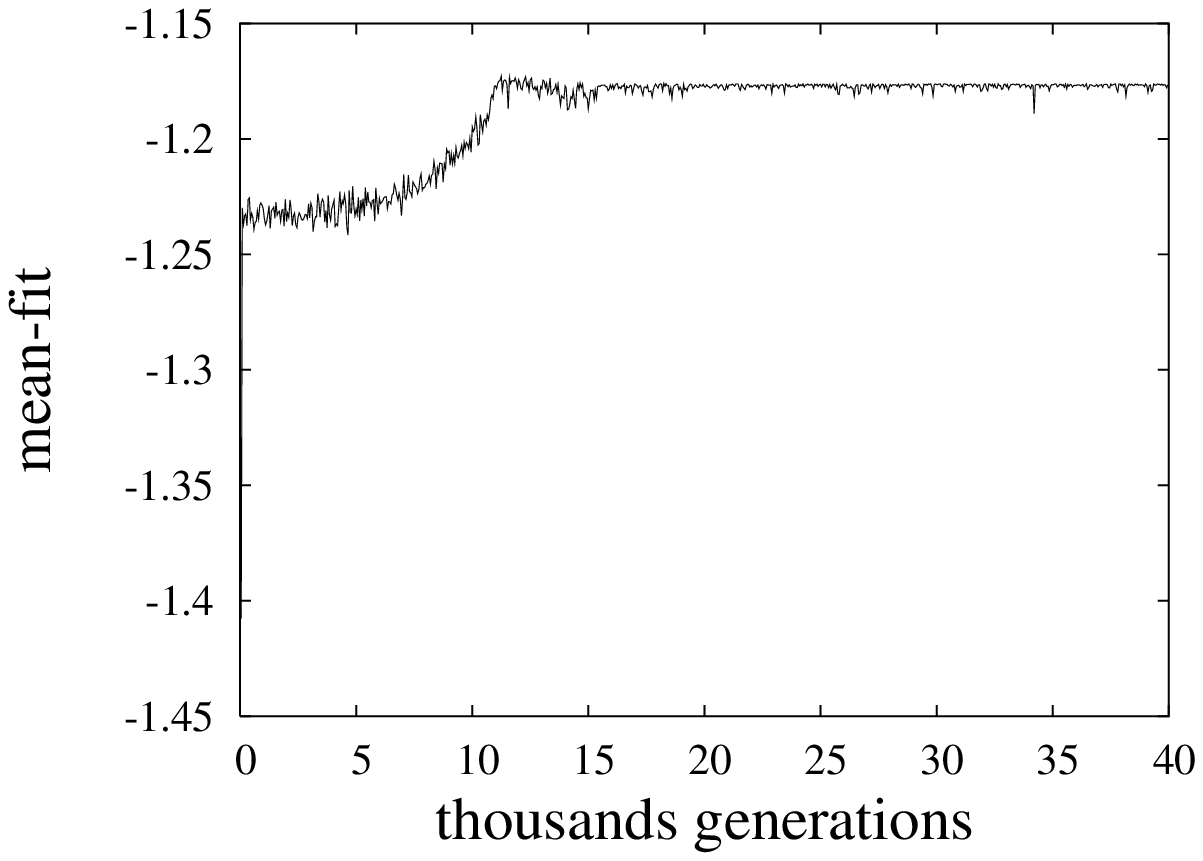} &
\includegraphics[scale=0.6]{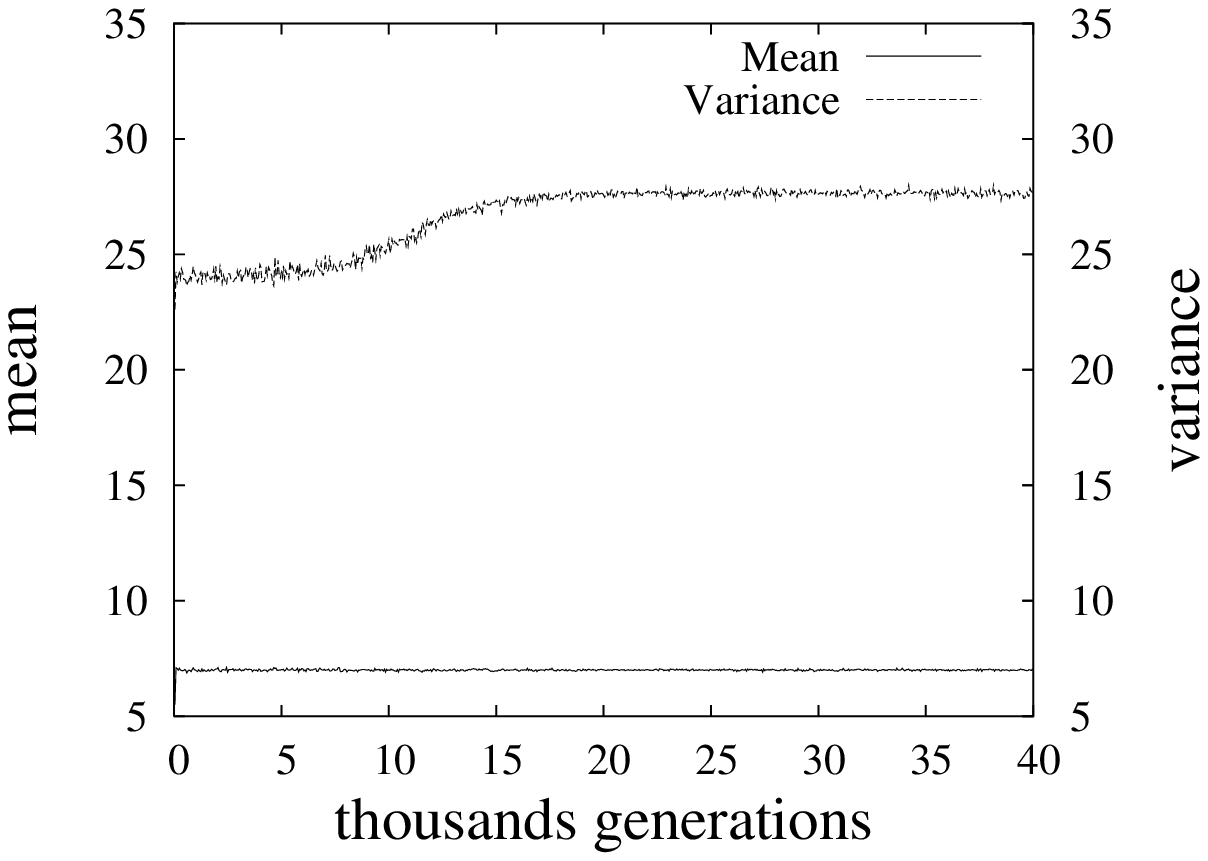}
\end{tabular}
  
 \caption{ Average fitness, variance and mean in a regime of strong assortativity ($\Delta = 1$) and high competition intensity but short competition range ($J=8$, $\alpha = 2$, $R = 2$). Annealing parameters: $\mu_0 = 10^{-1}$, $\mu_{\infty} = 10^{-6}$, $\tau = 10000$, $\delta = 3000$. Total evolution time: 40000 generations. }
   
\label{fitness-variance_4C}
\end{center}
\end{figure}

 Finally, it is important to compare the result of the simulation with $J=8$, $\alpha =2$, $R=2$ with that of the experiment with $J=1$, $\alpha =2$, $R=6$ portrayed in Figures~\ref{fig:4B} and~\ref{fitness-variance_4B}. In both cases the  competition is very strong, but when the  competition strength is due to the radius of competition, no more than two species can coexist and they are located at the opposite ends of the phenotypic space (at least when $R=6$); conversely if competition strength is due to the intensity parameter $J$, then four species will coexist (in the case $J=8$) so as to minimize intraspecific competition. Figure~\ref{fitness-variance_4C} shows the plots of mean fitness, variance and mean in the run with $J=8$.

\paragraph{Interplay between mating range and competition.}

 So far we have shown that a high assortativity level
alone ($\Delta = 1$, $\Delta = 0$) is sufficient to induce speciation, but the two newborn species cannot coexist for very long as one of them will be eradicated by random fluctuations. We then showed that the stable coexistence of the new species can be ensured by a high competition level, and in particular we showed the different effects of an increase in competition intensity and an increase in competition range. We now explore what happens if the mating range $\Delta$ is very large. In Section~\ref{moderate-assort} we showed that in the  absence of competition the  speciation is not possible, not even in a transient way, if we set a large mating range $\Delta = 4$. In the next simulation we show that speciation is indeed possible also in the case $\Delta = 4$ if we set a sufficiently strong competition: $J = 2$, $\alpha = 10$, $R = 4$. For the sake of comparison, recall that with $\Delta = 1$ a much weaker competition was sufficient: $J=1$, $\alpha = 2$, $R=2$ (see Figure~\ref{fig:4A} and Figure~\ref{fitness-variance_4A}). The results of the simulations, show that the final distribution is composed by two delta-peaks at the opposite ends of the phenotypic space \footnote{The first peak is typically placed at $x=1$ or $x=2$ or $x=3$ or $x=4$ and the second one at $x=11$, $x=12$, $x=14$, and less frequently also at $x=9$ and $x=14$.}.

\begin{figure}[ht]
\begin{center}  
\begin{tabular}{cc}
\includegraphics[scale=0.6]{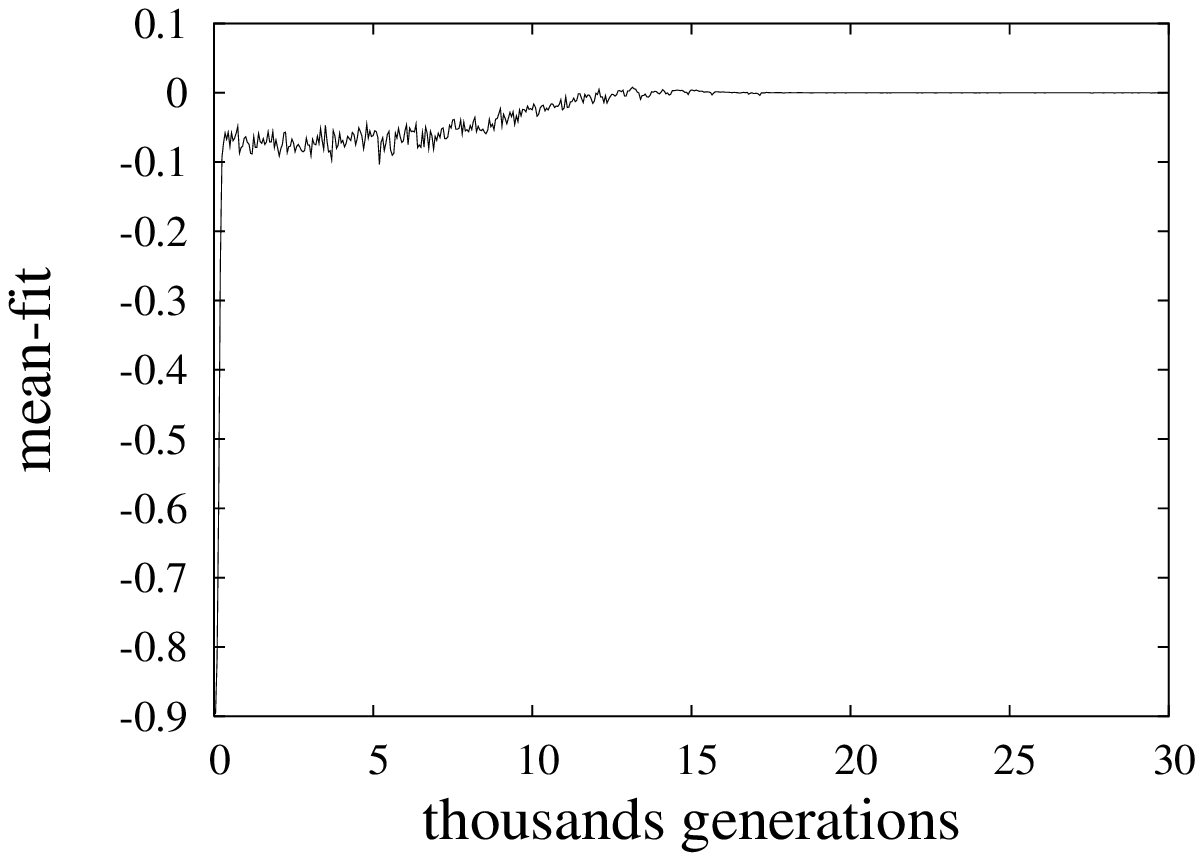} &
\includegraphics[scale=0.6]{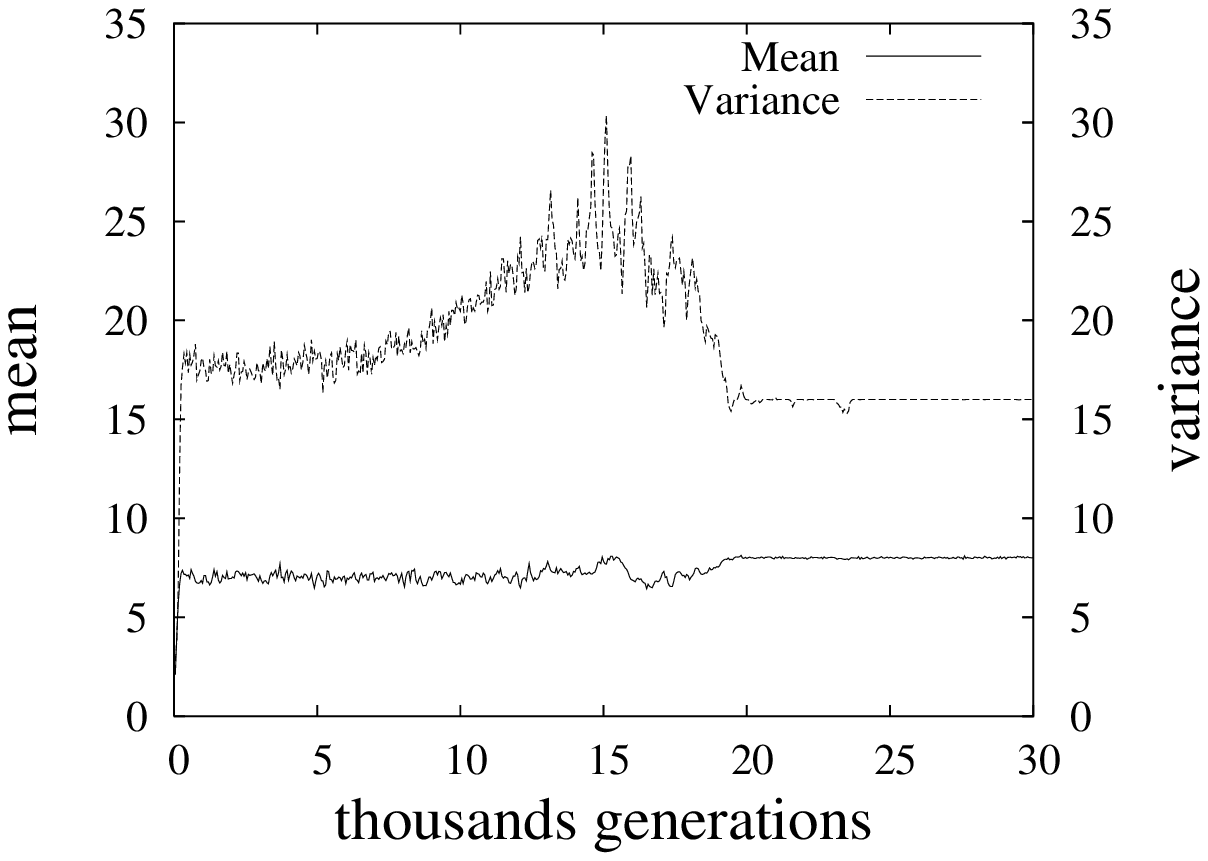}
\end{tabular}
   
\caption{ Average fitness, variance and mean in a regime of weak assortativity ($\Delta = 4$), high competition intensity and long competition range ($J=2$, $\alpha = 10$, $R = 4$). Annealing parameters: $\mu_0 = 10^{-1}$, $\mu_{\infty} = 10^{-6}$, $\tau = 10000$, $\delta = 3000$. Total evolution time: 30000 generations. }
     
\label{fitness-variance_4D}
\end{center}
\end{figure}

The plots of mean fitness, variance and mean can be easily interpreted. Due to the high competition level the initial distribution (in our case a delta-peak in $x=0$) splits within a few hundreds of generations in two bell-shaped distributions, the first one covering phenotypes $x=0$ to $x=6$ and the second one spanning the phenotypes $x=8$ to $x=14$. The first distribution has got the highest frequencies on phenotypes $x=2$, $x=3$, $x=4$, while the second one has got its maximum on $x=10$, $x=11$, $x=12$. The width of the two distributions is due to the high mutation rate and it does not change until $T \cong \tau$. After that, when the mutation rate decreases the two distributions are narrowed. The first phenotypes whose frequency decreases in both distributions are those placed towards the center of the phenotypic space because of the very high competition level. Later on, because of recombination, there is a decrease in frequency also for the phenotypes of both distributions placed towards the ends of the phenotypic space. This evolutionary dynamics is reflected in the plot of mean fitness. The mean fitness increases abruptly at the beginning of the run because the formation of two bell-shaped distributions relieves the competition pressure. When the two distributions narrow down and become delta-peaks, the average fitness further increases only to a small extent because the decrease in interspecific competition is almost completely compensated for by the increase in intraspecific competition. Finally, it should be noticed that the high level of competition is also reflected by the fact that the average fitness $\bar{H}$ is always negative, while in Figure~\ref{fitness-variance_4A} it was always positive. The plot of variance is also very interesting. After an abrupt increase at the beginning of the run, the variance oscillates around a constant value until $T \cong \tau$. After that the variance first increases and then decreases again. This is a consequence of the fact that the two distribution do not narrow down on a symmetric way  but they first lower their frequencies in the regions pointing towards the center of the phenotypic space and only later they reduce their frequencies in the opposite end of the distribution. Figure~\ref{fitness-variance_4D} shows the plots of mean fitness, variance and mean that we have just commented.

If we further increase the mating range $\Delta = 6$, the competition level necessary to induce speciation becomes even higher: $J=10$, $\alpha = 10$, $R = 6$. The evolutionary dynamics is the same as described in the case $\Delta = 4$ and the final distribution is made up of two delta peaks located in the extremes of the phenotype space \footnote{ One in $x=1$, $x=2$, $x=3$ or $x=4$ and the second one in $x=11$, $x=12$, $x=13$.}.
The plot of mean fitness shows values that are even more negative than in the simulation with $\Delta = 4$. Apart for this, there are no significant differences: the mean fitness increases abruptly at the beginning of the run and then undergoes a second, very modest increase after $T = \tau$. The variance, conversely after $T = \tau$ first shows a very little increase and then undergoes a more substantial decrease. This happens because, due to the very high competition pressure, the two bell-shaped distributions that appear at the beginning of the simulation, become asymmetrical long before $T = \tau$ showing a decrease in frequency of the phenotypes pointing towards the center of the phenotypic space. As a consequence, after $T = \tau$ there will be basically only the decrease in frequency of the phenotypes pointing towards the ends of the phenotypic space, hence the decrease in variance of the overall distribution. These patterns  are shown in Figure~\ref{fitness-variance_4D}.

\begin{figure}[ht]
\begin{center}     
\begin{tabular}{cc} 
\includegraphics[scale=0.6]{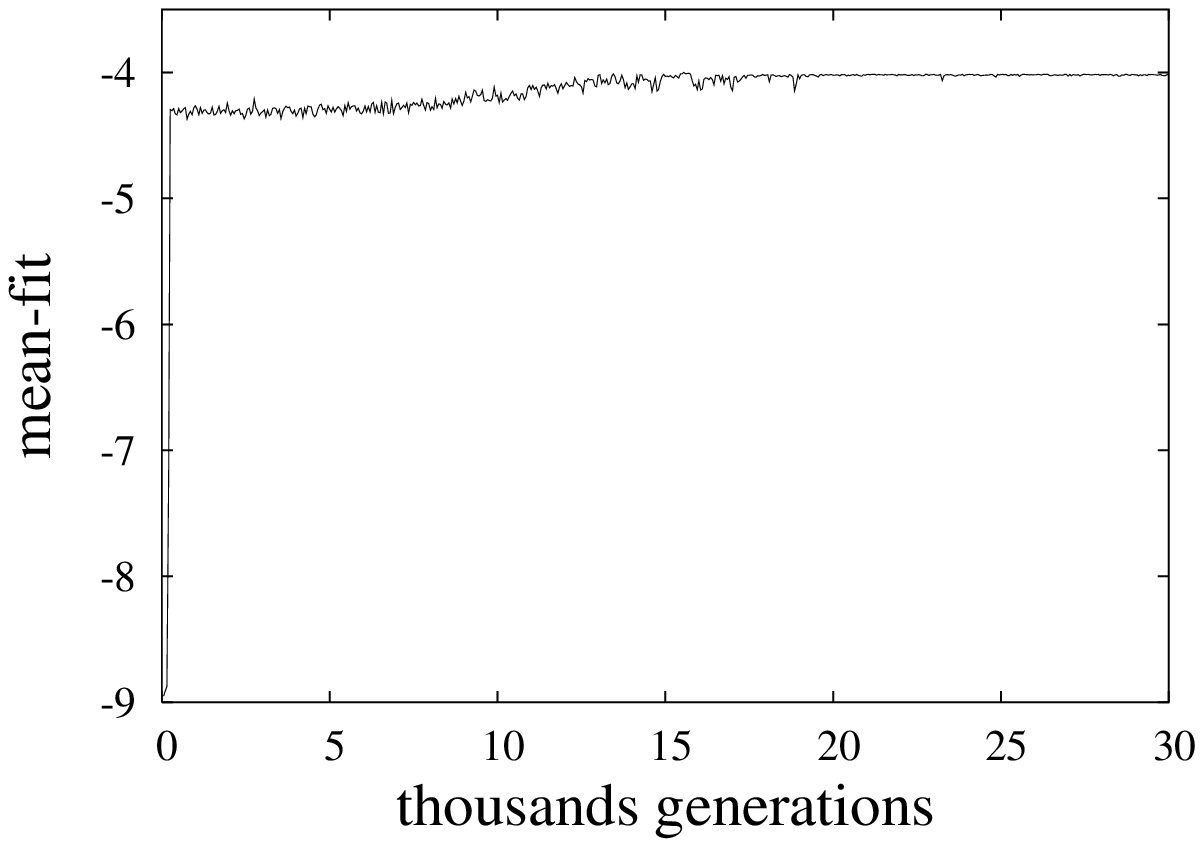} &
\includegraphics[scale=0.6]{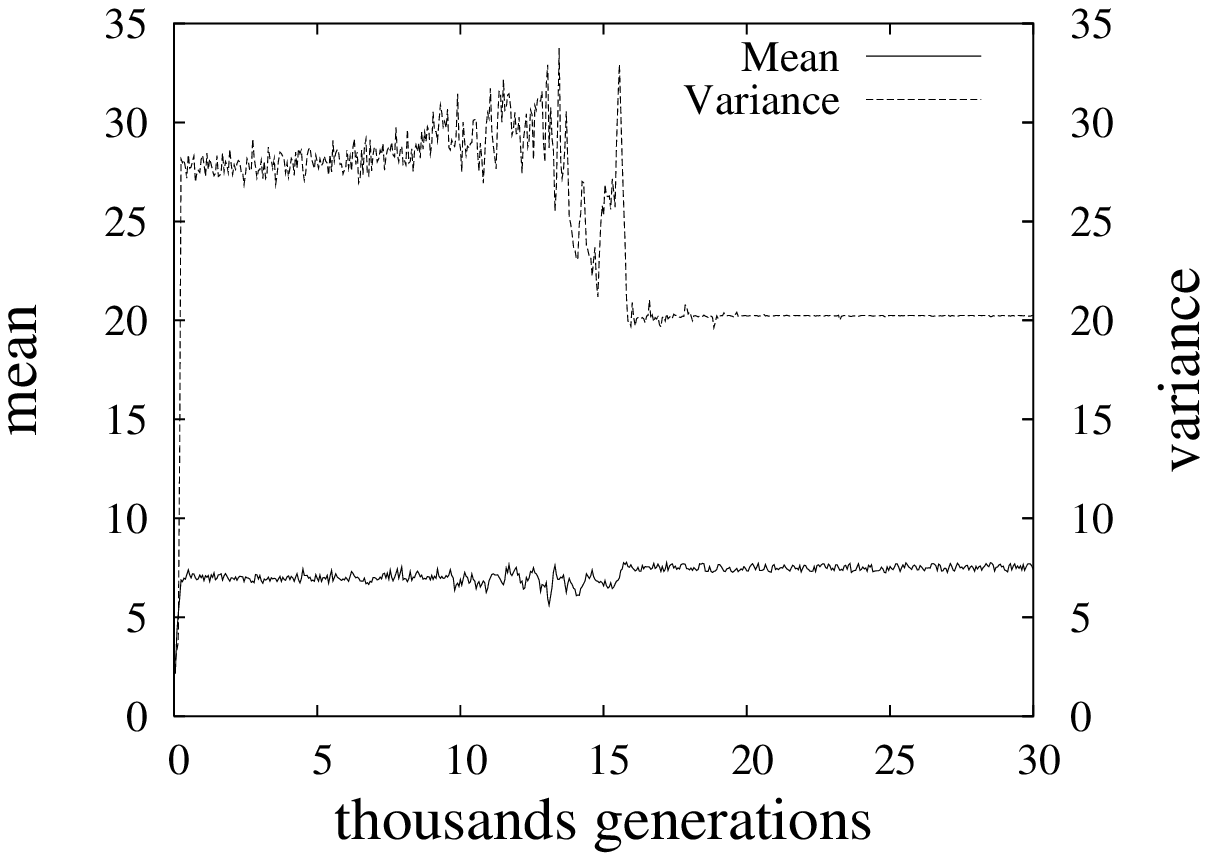}
\end{tabular} 
         
\caption{ Average fitness, variance and mean in a regime of very weak assortativity ($\Delta = 6$), high competition intensity and long competition range ($J=10$, $\alpha = 10$, $R = 6$). Annealing parameters: $\mu_0 = 10^{-1}$, $\mu_{\infty} = 10^{-6}$, $\tau = 10000$, $\delta = 3000$. Total evolution time: 30000 generations. } 
	       
\label{fitness-variance_4E}
\end{center}
\end{figure}

As a conclusion, our simulations in a flat static fitness landscape show that competition acts in two ways:

\begin{enumerate}

\item Competition stabilizes incipient species generated thanks to a high assortativity regime.

\item Competition induces speciation events that would be otherwise forbidden when the mating range is very large.

\end{enumerate}

Our simulations also show that competition and assortativity act in a synergistic way as very high levels of competition allow speciation even when assortativity is low; conversely a regime of very high assortativity makes speciation possible even in the face of a weak competition.

Finally, we have provided evidence that two dimensions of competition do exist: the competition intensity that induces the formation of a large number of species located at roughly equal distance in the phenotypic space so as to minimize competition pressure, and the competition range that induces the formation of a small number of species placed at maximal distance from each other.

\subsection{Steep static fitness landscape.}

\subsubsection{Absence of competition.}

In our model we assume that the 0-allele represents the wild-type
while the 1-allele is the least deleterious mutant. This entails that
the larger the number of 1-alleles in a genotype, the lower the
fitness.  This is why a particularly interesting case to investigate
is that of a monotonic decreasing static fitness.

\begin{figure}[ht!]
\begin{center}
\includegraphics[scale=0.7]{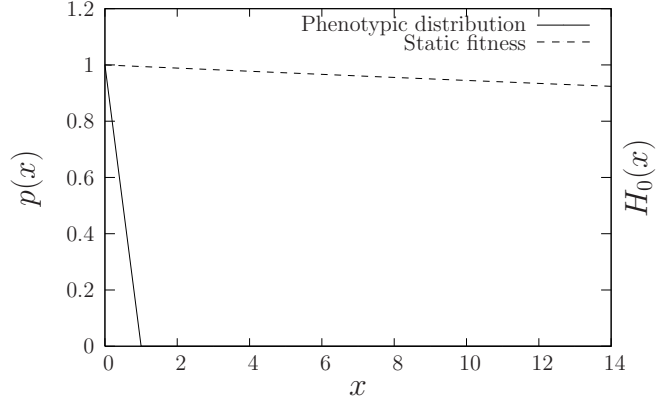}

\caption{The final distribution (generation 30000) obtained in absence of  competition  ($J = 0$) and with maximal assortativity ($\Delta = 0$) in the case of steep static fitness landscape ($\beta = 1$, $\Gamma = 150$) . Annealing parameters: $\mu_0 = (10)^{-1}$, $\mu_{\infty} = (10)^{-6}$, $\tau = 10000$, $\delta = 3000$. Total evolution time 30000 }
\label{fig:6}
\end{center}
\end{figure}

In the absence of competition the asymptotic distribution is a sharp
peak on the master sequence $x=0$. In this case, the whole population crowds on the phenotype with the
highest fitness level. The process is rather straightforward: at the beginning of the simulated annealing the initial frequency distribution is rapidly transformed in a distribution with a high peak in $x=0$ and a long tail of mutants at very low frequencies spanning over the entire phenotypic space. For $T> \tau$ the mutation rate decreases and the tail of mutants disappears so that the stationary distribution is represented by a delta peak in $x=0$. In Figure~\ref{fig:6} we show the stationary distribution in a run with: $\beta = 1$, $\Gamma = 150$, $J = 0$, $\Delta = 0$.

The behavior of the mean, variance and mean fitness of the distribution is straightforward and completely dominated by the mutation rate and the slope of the fitness landscape. As an example, when $p=0$ the initial distribution is a delta peak in $x=0$ so that the mean fitness equals one and the variance and mean equal zero. Due to the high mutation rate at the beginning of the simulation, however, a long tail of mutants appear in regions with low static fitness so that the mean fitness drops abruptly while the variance and mean of the distribution do increase. Finally, after $T = \tau$ the mutation rate decreases, the tail of mutants disappears and the distribution is turned again in a delta-peak in $x=0$. As a consequence, the mean and variance of the distribution vanish again while the mean fitness increases again to unity.

\subsubsection{Stabilizing effects of competition.}

Once again, competition plays a key role as a powerful stabilizing
force. When mating is random, the only effect of an extremely high
competition is to induce the formation of a wide and flat frequency
distribution spanning over all the phenotypic space, but no distinct
species will appear.

In conditions of high assortativity $\Delta=3$, a moderate level of
competition ($J=2$) is sufficient to induce the appearance of two
asymmetric sharp peaks. The tallest peak appears on the master
sequence because this is the position of the phenotypic space with
the highest static fitness; the other peak, at the opposite end of
the phenotypic space, is smaller because it is near the static fitness
minimum but it won't go extinct because, being very far from the
first peak, it enjoys the lowest possible interspecific competition.

\begin{figure}[ht!]

\begin{tabular}{ccc}
\hspace{-2 cm} & \includegraphics[width=8cm,height=6cm]{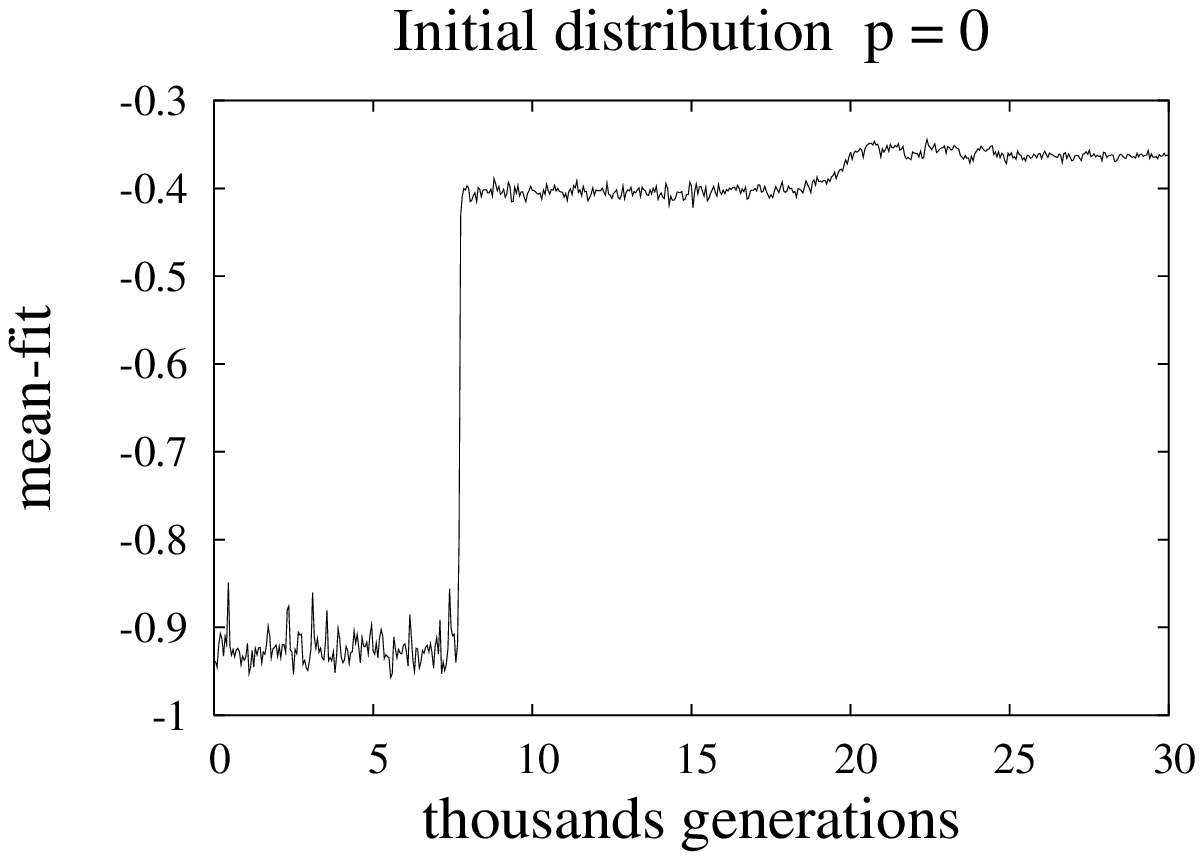} &
\includegraphics[width=8cm,height=6cm]{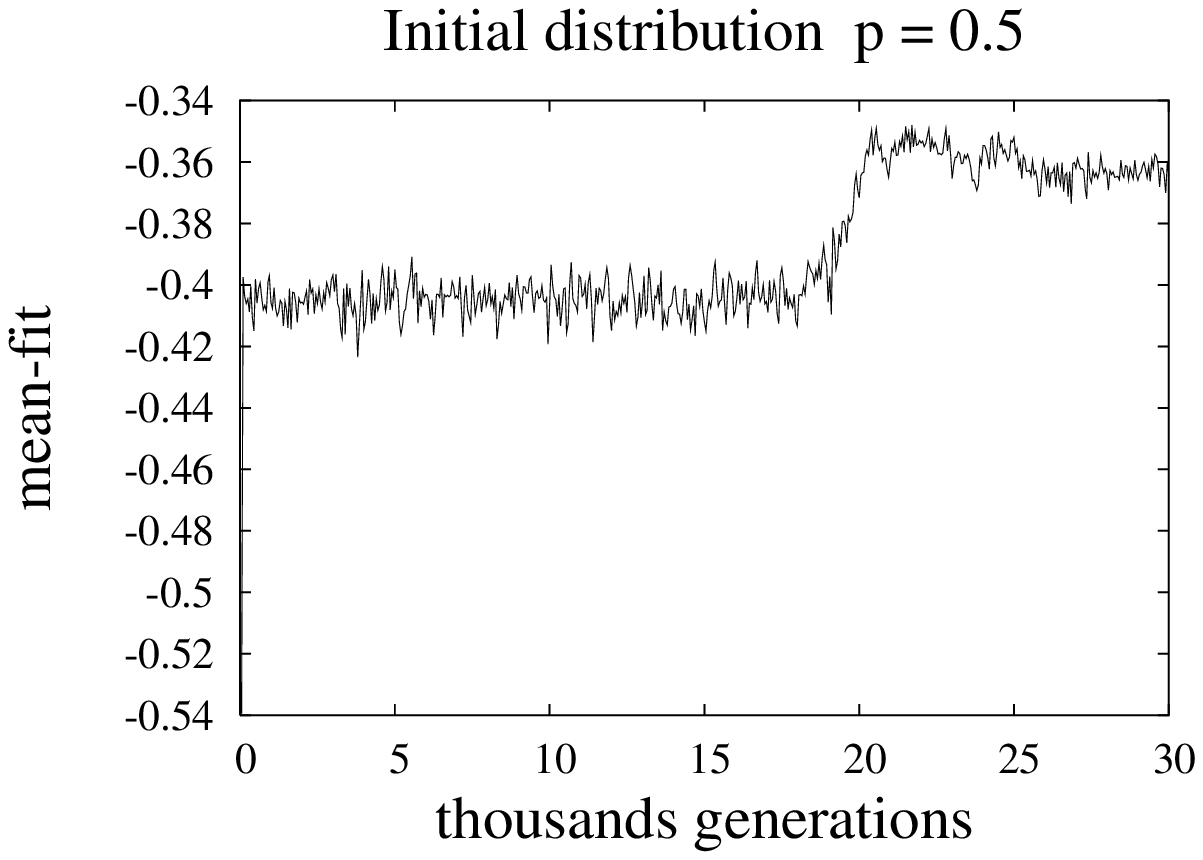} \\
\hspace{-2 cm} & \includegraphics[width=8cm,height=6cm]{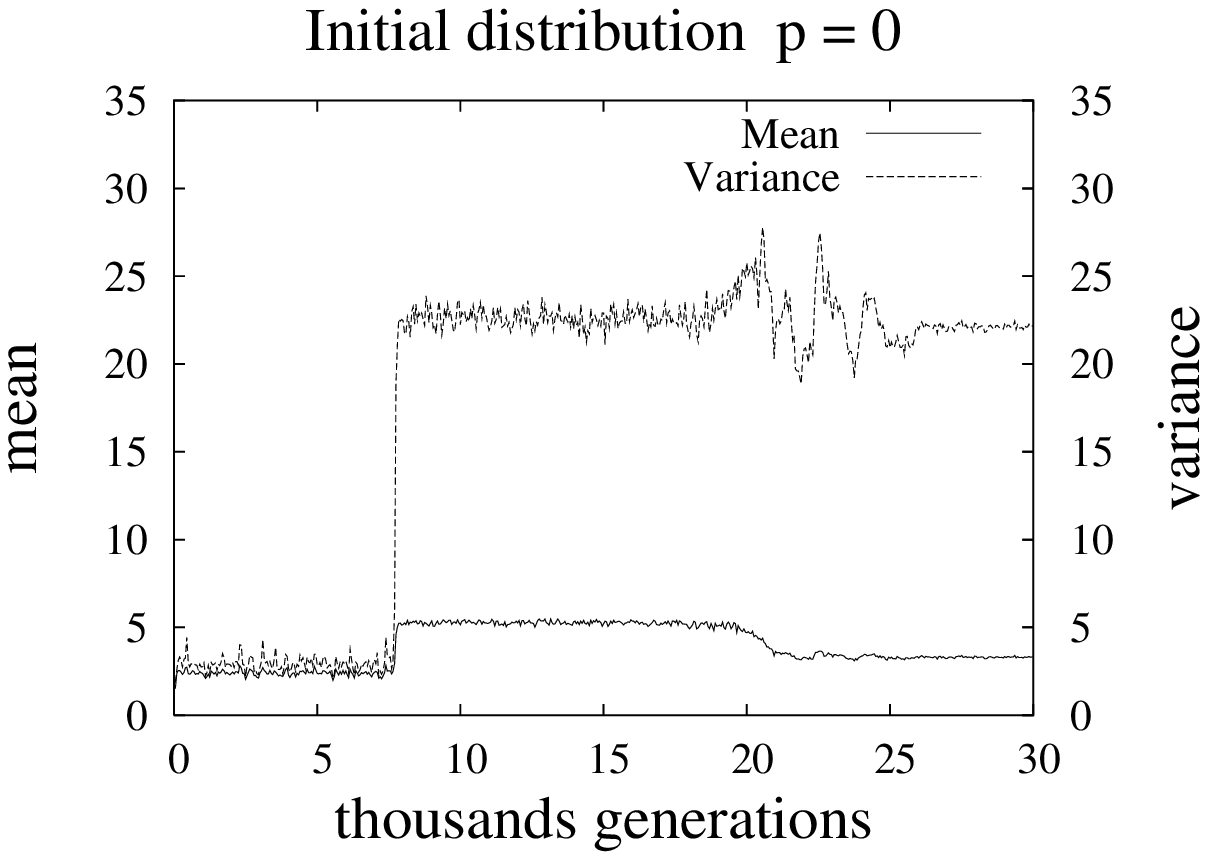} &
\includegraphics[width=8cm,height=6cm]{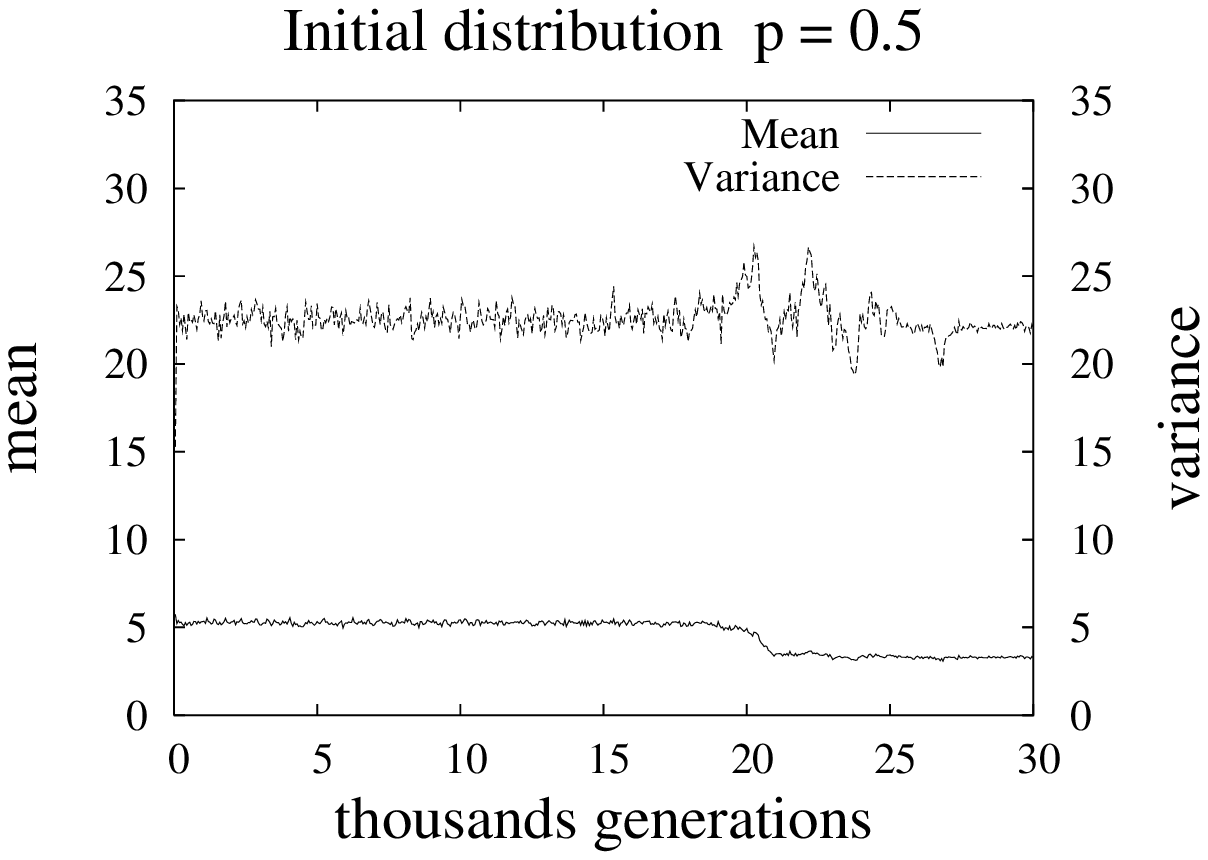}
\end{tabular}

\caption{plots of mean fitness, variance and mean for $p=0$ and $p=0.5$  obtained with moderate  competition  ($J = 2$) and  assortativity ($\Delta = 3$) in the case of steep static fitness landscape ($\beta = 1$, $\Gamma = 10$) . Annealing parameters: $\mu_0 = (10)^{-1}$, $\mu_{\infty} = (10)^{-6}$, $\tau = 20000$, $\delta = 1000$. Total evolution time 30000 }

\label{fitness-variance_7A}

\end{figure}

As an example, we describe the dynamic behavior in a simulation with the following parameter setting: $\beta = 1$, $\Gamma = 10$, $J = 2$, $\alpha = 2$, $R = 4$, $\Delta = 3$. It is necessary to distinguish the case $p  = 0.5$ and $p=0$. If $p=0.5$ the initial distribution quickly splits in two bell-shaped distributions, one near $x=0$ and one in the neighborhood of $x=14$. The distribution near $x=0$ is asymmetrical having higher frequencies on the phenotypes with higher static fitness; conversely, the distribution near $x=14$ is flat and wide because of the low values of the static fitness  in that region. The appearance of the two distributions at the opposite ends of the phenotypic space is related to a steep increase in mean fitness because spreading the population over several phenotypes relieves competition. For $T> \tau$ the mutation rate decreases and the two bell-shaped distributions narrow down as a result of recombination and the second one moves closer to the first one following the static fitness gradient so that both the variance and the mean are decreased while the mean fitness increases to a small extent. On the contrary, if $p=0$, only an asymmetrical bell-shaped distribution near $x=0$ appears at the beginning of the simulation. A second bell-shaped distribution appears near $x=L$ only much later during the simulation when a colony of mutants large enough to resist to random fluctuations will be created in that region of the phenotypic space. As a consequence, when $p=0$ the abrupt increase in mean fitness will be delayed as compared to the case $p>0$ and the same applies to the variance and the mean. For $T > \tau$ however, the pattern is the same as for $p>0$ and in the final distribution we always find a tall delta peak in $x=0$ and a second smaller  peak near the opposite end of the phenotypic space  \footnote{Usually in position $x=10$ but less often also in $x=9$ and $x=11$.}.
Figure~\ref{fitness-variance_7A} shows the plots of mean fitness, variance and mean in the cases $p=0$ and $p=0.5$.

\begin{figure}[ht!]
\begin{center}
\begin{tabular}{cc}
\includegraphics[scale=0.6]{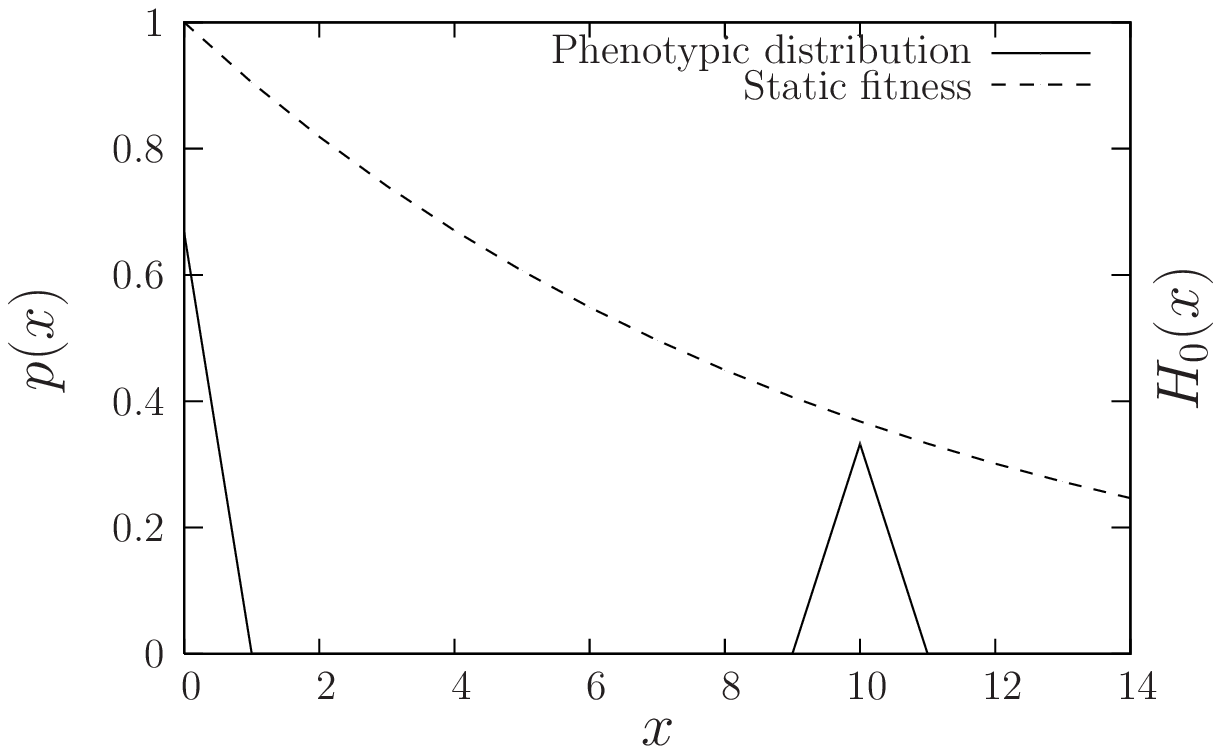}
\includegraphics[scale=0.6]{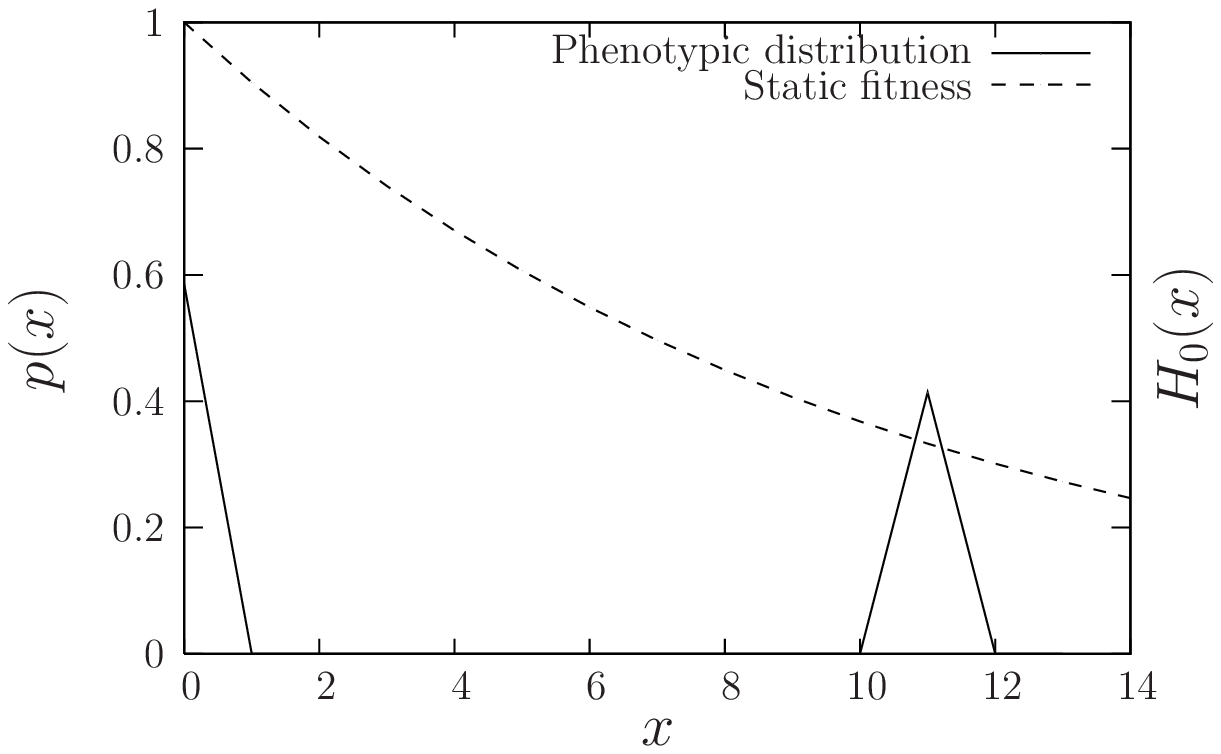}
\end{tabular}

\caption{The final distribution (generation 40000) obtained with  competition  intensity $J = 2$ (left panel) and $J=4$ (right panel) for a moderate  assortativity $\Delta = 3$ in the case of steep static fitness landscape ($\beta = 1$, $\Gamma = 10$) . Annealing parameters: $\mu_0 = (10)^{-1}$, $\mu_{\infty} = (10)^{-6}$, $\tau = 20000$, $\delta = 1000$. Total evolution time 40000 }
\label{fig:7A-7B1}
\end{center}   
\end{figure}

As mentioned above, when competition is weak ($J=2$) the final distribution is formed by two asymmetric delta peaks. The peak in $x=0$ is populated by $65 \%$ of the population while the peak in  $x = 10$ accounts for the  remaining $35 \%$ of the organisms. When the competition intensity is increased to $J = 4$, the final distribution still shows two peaks, but these are almost symmetrical: the delta peak in $x=0$ is populated by about $55 \%$ of the individuals, while the second peak which is usually a delta peak in $x=11$ but sometimes also a peak spanning over phenotypes $x=10$ and $x =11$ or phenotypes $x=11$ and $x=12$, comprises the remaining $45 \%$. Figure~\ref{fig:7A-7B1} shows the typical final distribution attained in these simulations.

\subsubsection{Competition-induced speciation.}

\begin{figure}[ht!]
\begin{center}
\includegraphics[scale=0.7]{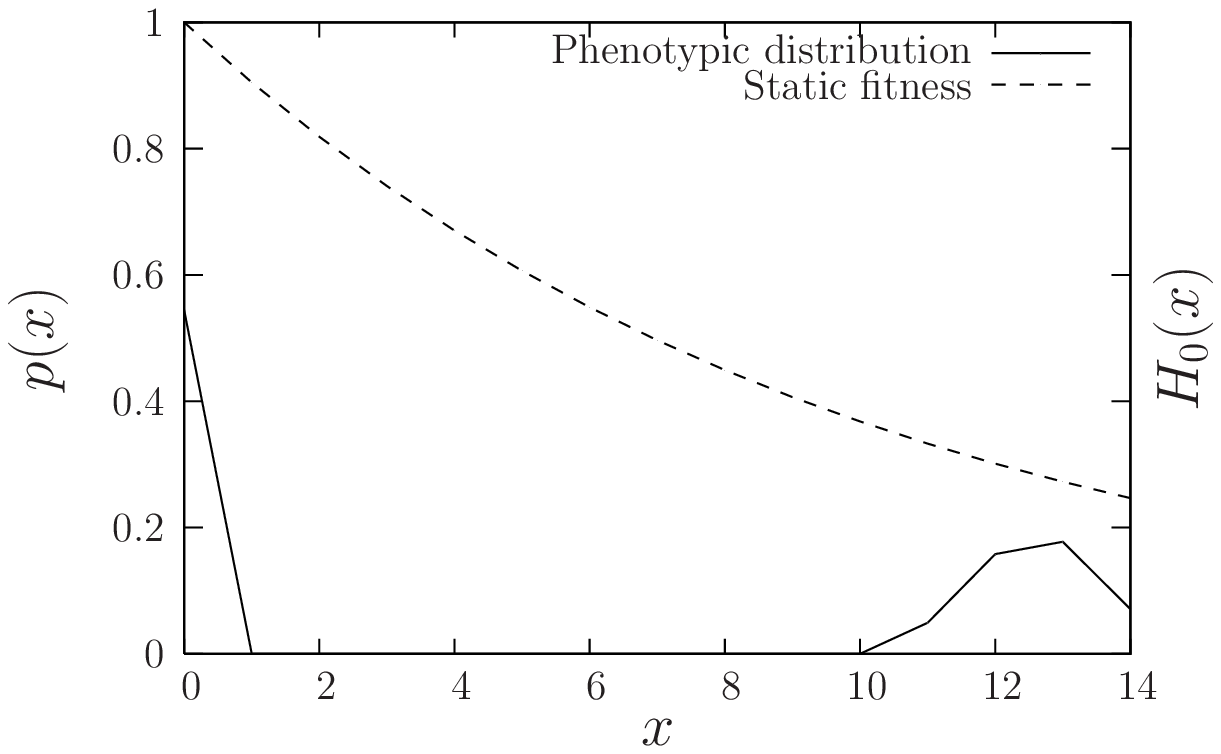}

\caption{The final distribution (generation 40000) obtained with  competition  intensity $J = 7$ and  assortativity $\Delta = 3$ in the case of steep static fitness landscape ($\beta = 1$, $\Gamma = 10$) . Annealing parameters: $\mu_0 = (10)^{-1}$, $\mu_{\infty} = (10)^{-6}$, $\tau = 20000$, $\delta = 1000$. Total evolution time 40000 }
\label{fig:7B2}
\end{center}
\end{figure}

If competition is increased up to $J=7$, the population splits in two
species with approximately the same frequency but whose distributions
show very different geometries. While the master sequence species
still exhibits a sharp peak distribution populated by about $55 \%$ of the population, the species near $x=L$
shows a wide and flat bell-shaped frequency distribution so as to
minimize the intraspecific competition \footnote{This second distribution usually extends over phenotypes $x=11$, $x=12$ and $x=13$ but sometimes also over phenotypes $x=10$ and/or $x=14$.}. The final distribution obtained in a simulation with $J=7$ and all the other parameters unchanged as compared to the previous example is shown in Figure~\ref{fig:7B2}.

\begin{figure}[ht!] 
\begin{center} 
\includegraphics[scale=0.7]{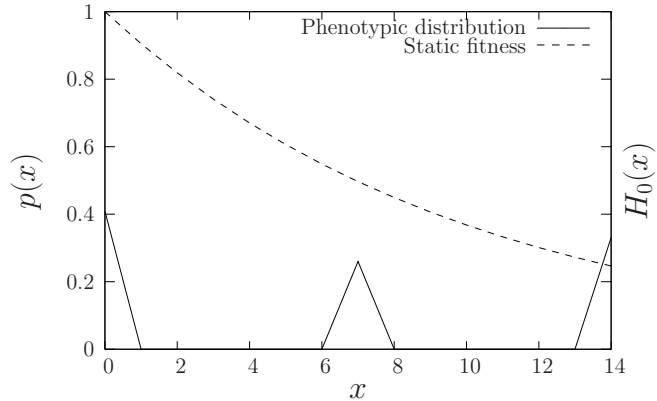}
  
\caption{The final distribution (generation 40000) obtained with  competition  intensity $J = 10$ and  assortativity $\Delta = 3$ in the case of steep static fitness landscape ($\beta = 1$, $\Gamma = 10$) . Annealing parameters: $\mu_0 = (10)^{-1}$, $\mu_{\infty} = (10)^{-6}$, $\tau = 20000$, $\delta = 1000$. Total evolution time 40000 } 
\label{fig:7B3} 
\end{center}   
\end{figure}

We will now investigate what happens if $J$ is increased to 10 while the other parameters are left unchanged. At the beginning of the simulation a frequency distribution appears, that spans the whole phenotypic space and shows maximal frequencies near $x=0$ and $x=14$; there is also a hump in the central region of the phenotypic space: these are the regions where competition is minimal. Moreover, the phenotypes close to $x=0$ and $x=14$ enjoy a very little dispersion of offsprings. The creation of such a distribution implies a first significant increase in mean fitness, variance and mean. Immediately after $T = \tau$ the mutation rate decreases and three delta-peaks appear in $x=0$, $x=7$ and $x=14$. This leads to a new, significant increase in mean fitness because, contrary to what happens in the cases $J=4$ and $J=7$, the effect of the increase in frequency of phenotype $x=0$ that enjoys the maximal static fitness, is not canceled out by the increase in intraspecific competition, this event being prevented by the formation of a delta-peak in intermediate position. Immediately after $T = \tau$ there is also a significant increase in variance  because there is a significant increase in frequency of phenotypes $x=0$ and $x=14$, that being very far from the mean, provide a large contribution to the variance, while in the same time there is a decrease in frequency of the phenotypes in intermediate position close to the mean. As usual, in the case $p=0.5$ the mean fitness and variance of the initial population is very large so that the initial increase of these parameters will be more limited. In Figure~\ref{fig:7B3} we show a typical example of final distribution when $J=10$.

\begin{figure}[ht!]

\begin{tabular}{ccc}
\hspace{-2 cm} & \includegraphics[width=8cm,height=6cm]{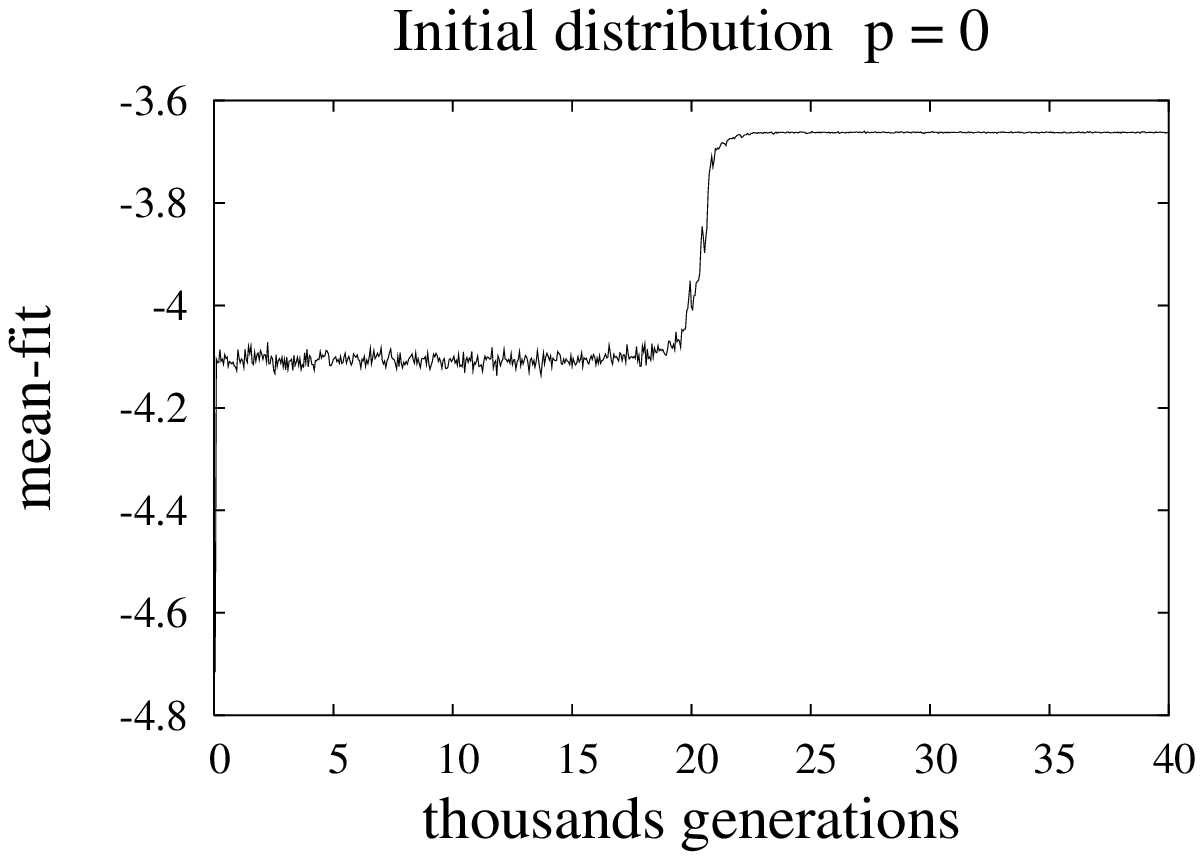} &
\includegraphics[width=8cm,height=6cm]{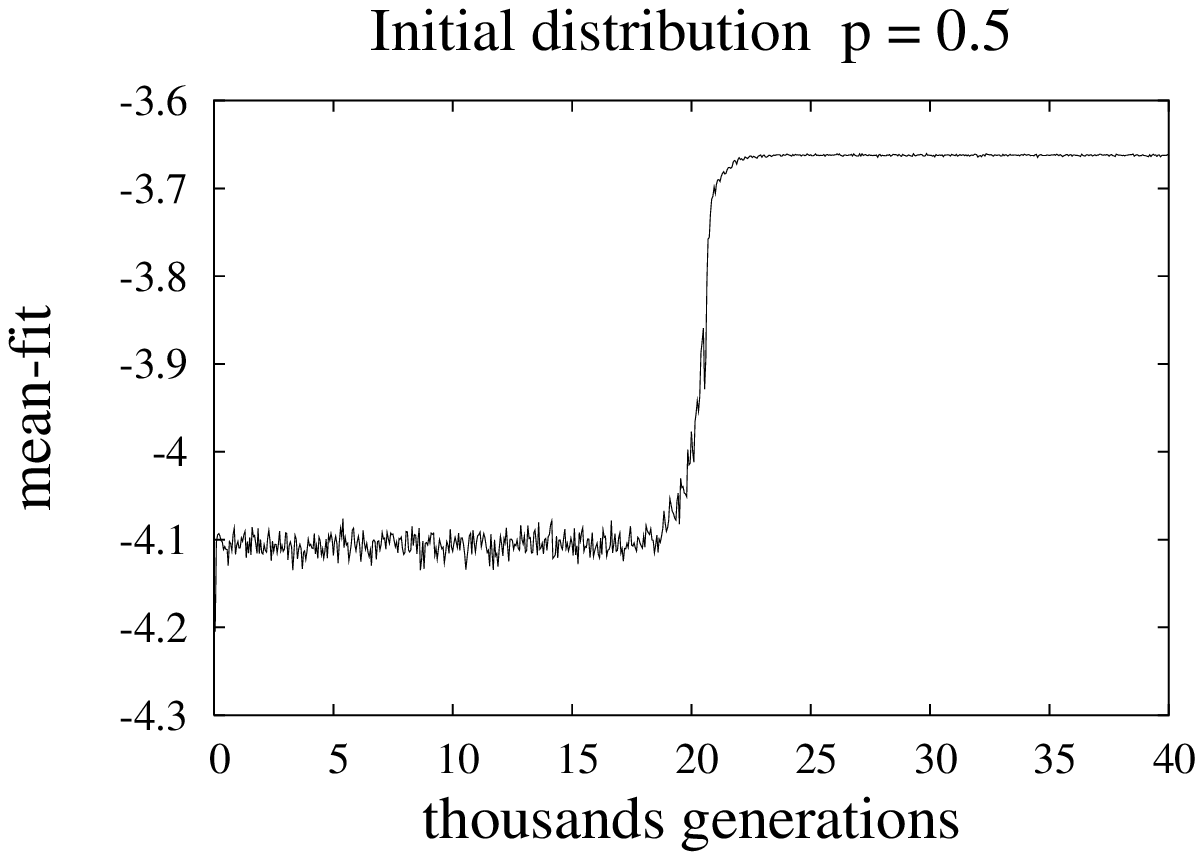} \\
\hspace{-2 cm} & \includegraphics[width=8cm,height=6cm]{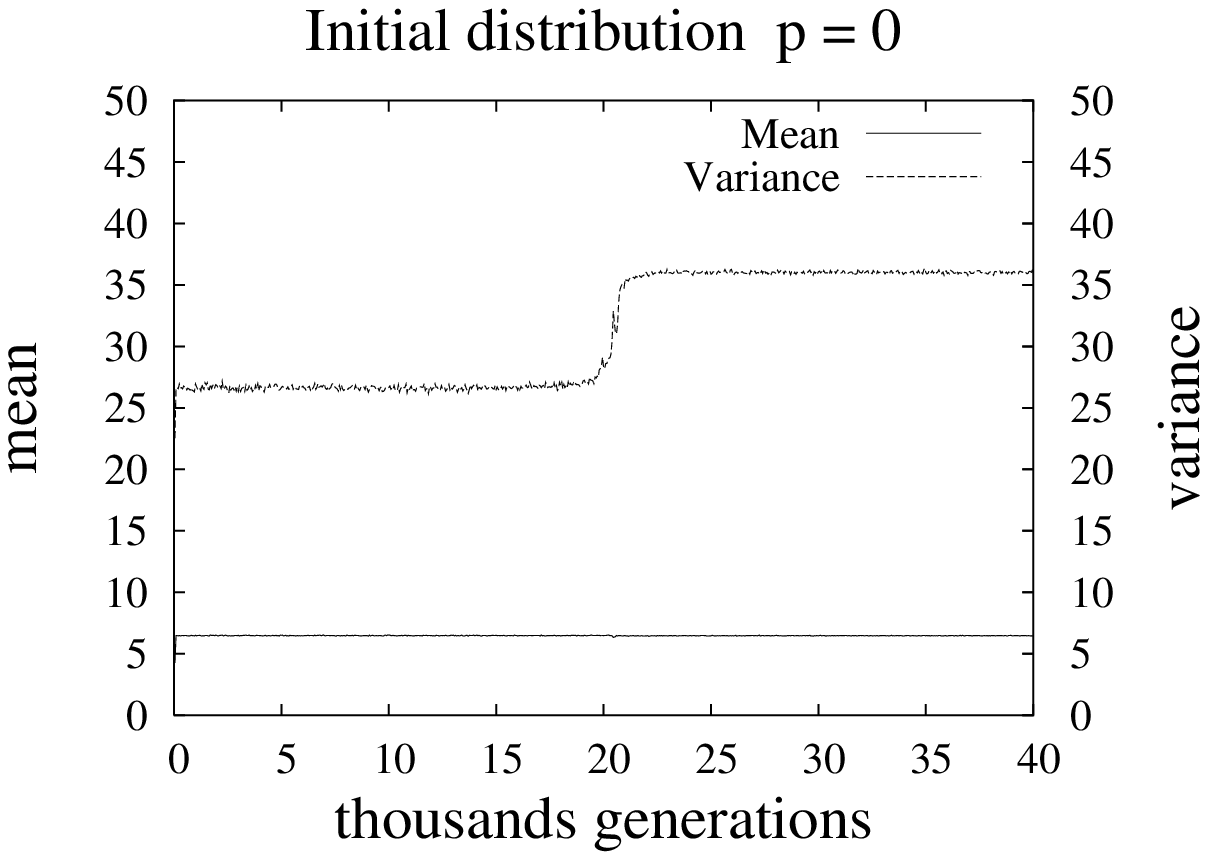} &
\includegraphics[width=8cm,height=6cm]{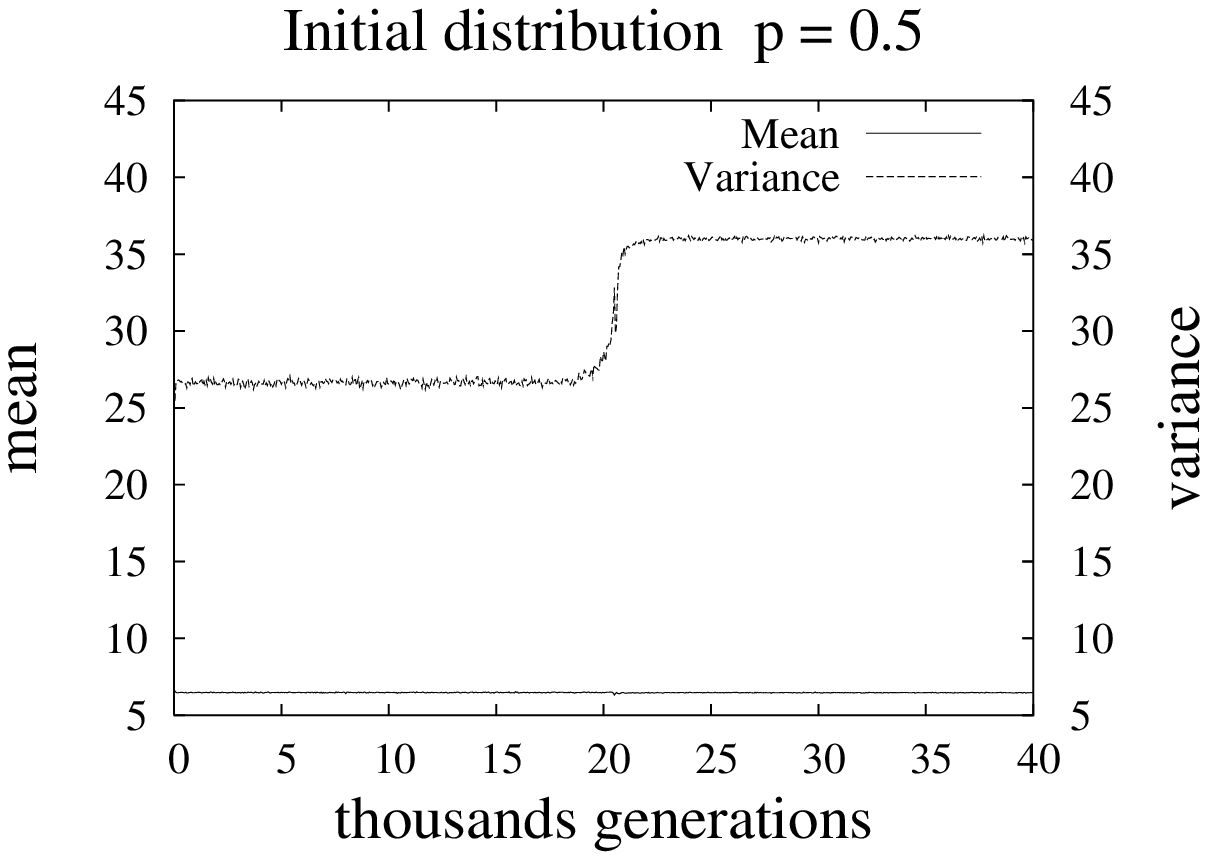}
\end{tabular}
 
\caption{Plots of mean fitness, variance and mean for $p=0$ and $p=0.5$  obtained with competition intensity  $J = 10$ and  assortativity $\Delta = 3$ in the case of steep static fitness landscape ($\beta = 1$, $\Gamma = 10$) . Annealing parameters: $\mu_0 = (10)^{-1}$, $\mu_{\infty} = (10)^{-6}$, $\tau = 20000$, $\delta = 1000$. Total evolution time 40000 }

\label{fitness-variance_7B3}
     
\end{figure}

In Figure~\ref{fitness-variance_7B3}, we also show the plots of mean fitness, variance, mean in the cases $p=0$ and $p=0.5$ when $J=10$.

\subsubsection{Interplay between competition and assortativity in a steep fitness landscape.}

\begin{figure}[ht!]
\begin{center}
\begin{tabular}{cc}
\includegraphics[scale=0.6]{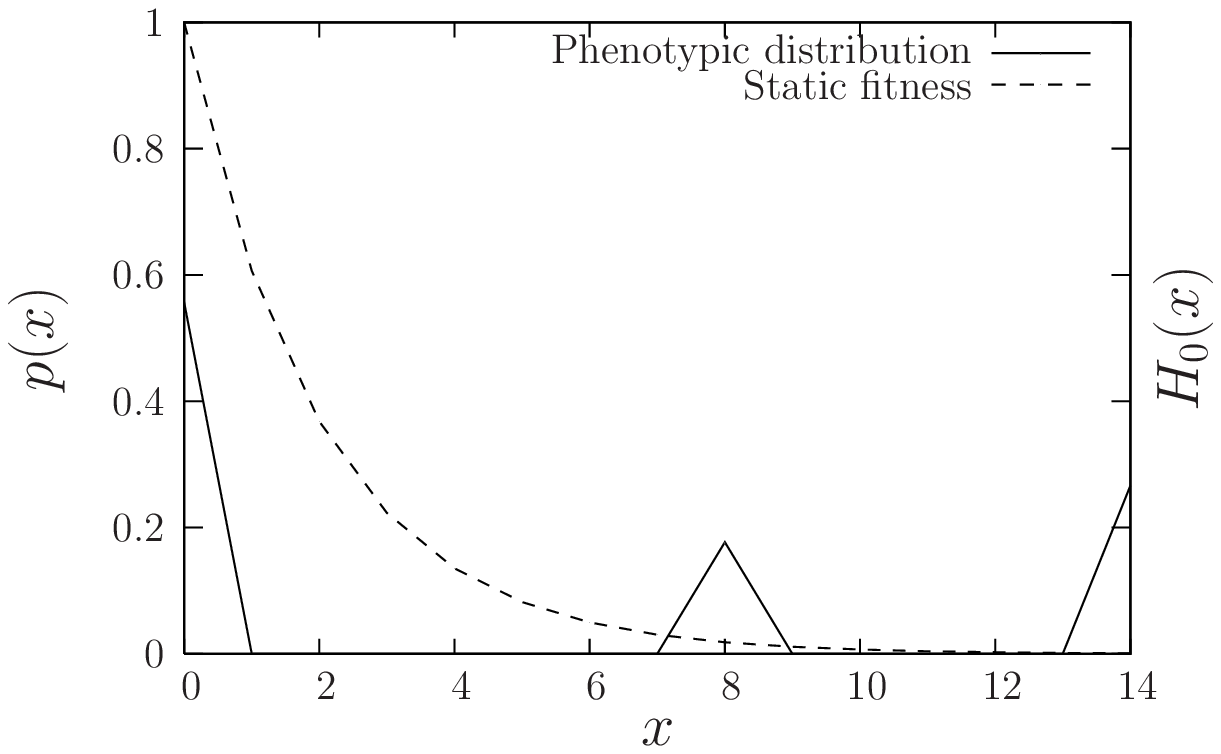} &
\includegraphics[scale=0.6]{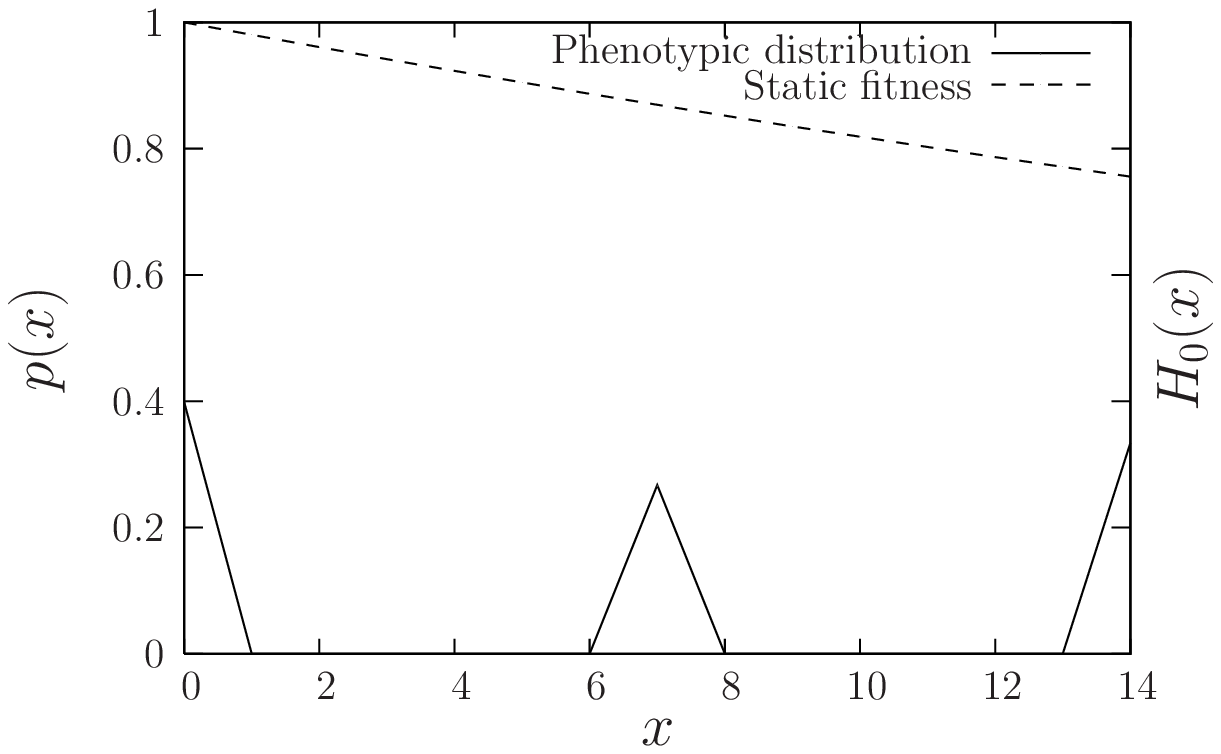}
\end{tabular}
\caption[scale=0.7]{The final distribution (generation 30000) obtained with  competition  intensity $J = 4$ and maximal  assortativity $\Delta = 0$ in the case of an extremely steep static fitness landscape ($\beta = 1$, $\Gamma = 2$) (left panel) and a  moderately steep one ($\beta = 1$, $\Gamma = 50$) (right panel)  . Annealing parameters: $\mu_0 = (10)^{-1}$, $\mu_{\infty} = (10)^{-6}$, $\tau = 20000$, $\delta = 1000$. Total evolution time 30000 }
\label{fig:8A-8B}
\end{center}
\end{figure}

Figure~\ref{fig:7B3} shows that a stationary distribution with three delta peaks can be obtained in the presence of moderate assortativity ($\Delta = 3$) if competition intensity is sufficiently high ($J=10$). We now show that this pattern can be obtained also with a much weaker competition ($J=4$) if assortativity is maximal ($\Delta = 0$). Maximal assortativity in fact, induces genotypical homogeneity in the strains of a population thus preventing dispersion of the offsprings. This is a very important condition for a peak in the middle of the phenotypic space  to resist eradication, as it experiences strong competition from both extreme strains. The symmetry of the stationary distribution depends on the steepness of the static fitness. When the static fitness is not so steep ($\Gamma = 50$) the three peaks are almost equal: $x=0$ ($40 \%$), $x=7$ ($30 \%$), $x=14$ ($30 \%$); conversely, when the fitness profile is very steep ($\Gamma = 2$) the peaks are very asymmetrical: $x=0$ ($50 \%$), $x=8$ ($20 \%$), $x=14$ ($30 \%$). The stationary distributions obtained with $\Gamma = 2$ and $\Gamma = 50$ are shown in Figure~\ref{fig:8A-8B}.

The plots of mean fitness, variance and mean only show quantitative differences from those shown in Figure~\ref{fitness-variance_7B3} and therefore they are not presented.

\subsection{Influence of  genome length.}

The simulations with both flat and steep static fitness landscape discussed so far, showed that the mating range $\Delta$ has got a strong influence on the evolutionary dynamics and on the stationary distribution. Small values of $\Delta$ in fact, make speciation possible in the presence of weak competition and sometimes even in the absence of competition. One however may argue that it makes sense to talk about a \emph{small} or \emph{large} mating range, only with reference to the genome length $L$ because this, on turn, determines the number of possible phenotypes and hence the size of the phenotypic space. In all the simulations discussed so far the genome length was set to $L = 14$. We now perform simulations with $L = 28$  to show that the final distribution is not affected by the genome length.

We begin with simulations with flat static fitness landscape. These simulations will have to be compared with those shown in Figures~\ref{fig:4A} and~\ref{fitness-variance_4A}. When the genome length is doubled from $L=14$ to $L=28$, the competition range $R$ and the mating range $\Delta$ must also be doubled from 2 to 4 and from 1 to 2 respectively. We also doubled the steepness parameter $\Gamma$ of the static fitness from 14 to 28 so as to achieve a flat landscape. 

\begin{figure}[ht!]
\begin{center}
\includegraphics[scale=0.7]{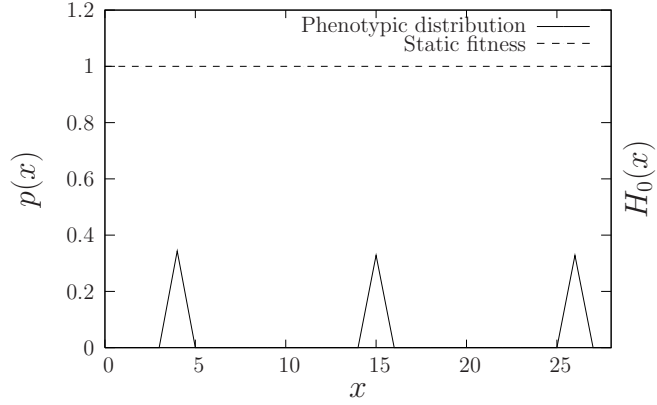}

\caption{Genome length $L = 28$.  The final distribution (generation 40000) obtained with  competition  intensity $J = 1$ and   assortativity $\Delta = 2$ in the case of flat fitness landscape ($\beta = 100$, $\Gamma = 28$) . Annealing parameters: $\mu_0 = (10)^{-1}$, $\mu_{\infty} = (10)^{-6}$, $\tau = 10000$, $\delta = 3000$. Total evolution time 40000 }
\label{fig:9}
\end{center}
\end{figure}

\begin{figure}[ht!]
\begin{center}
\begin{tabular}{cc}
\includegraphics[scale=0.6]{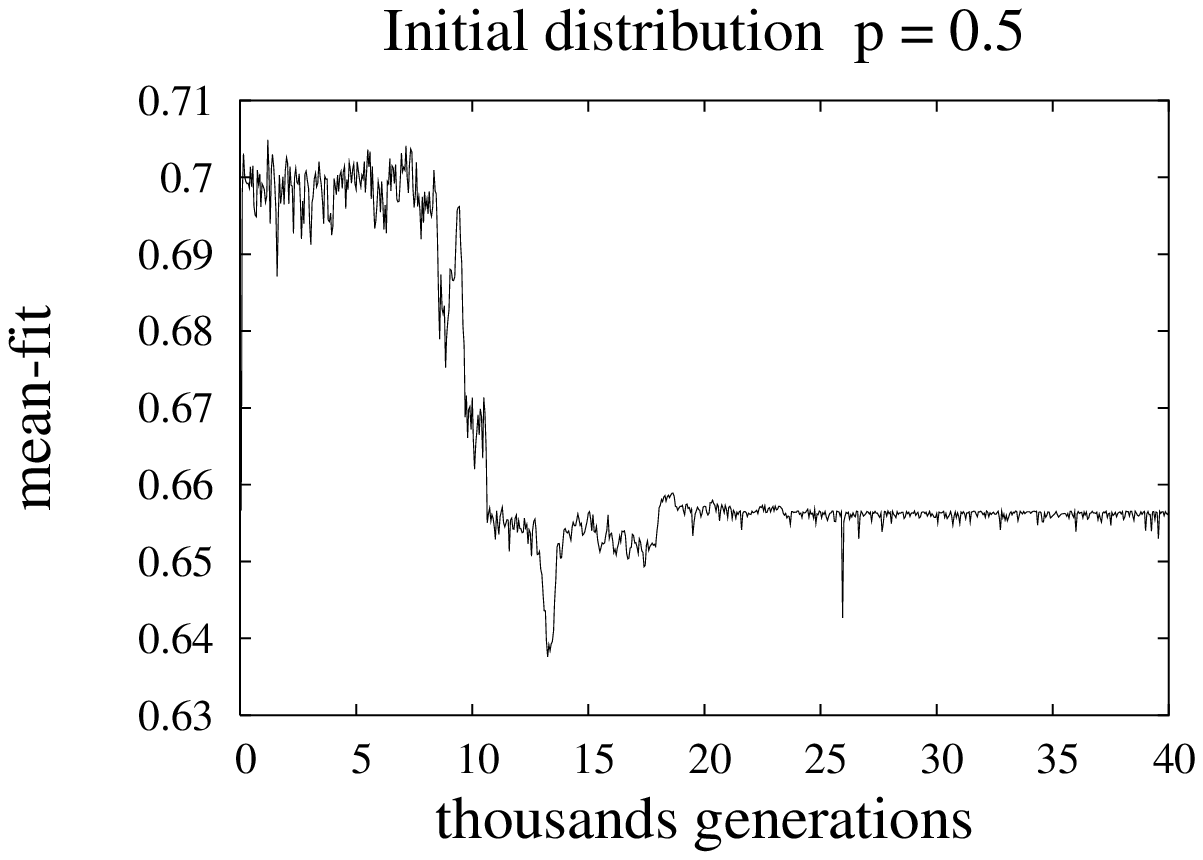} &
\includegraphics[scale=0.6]{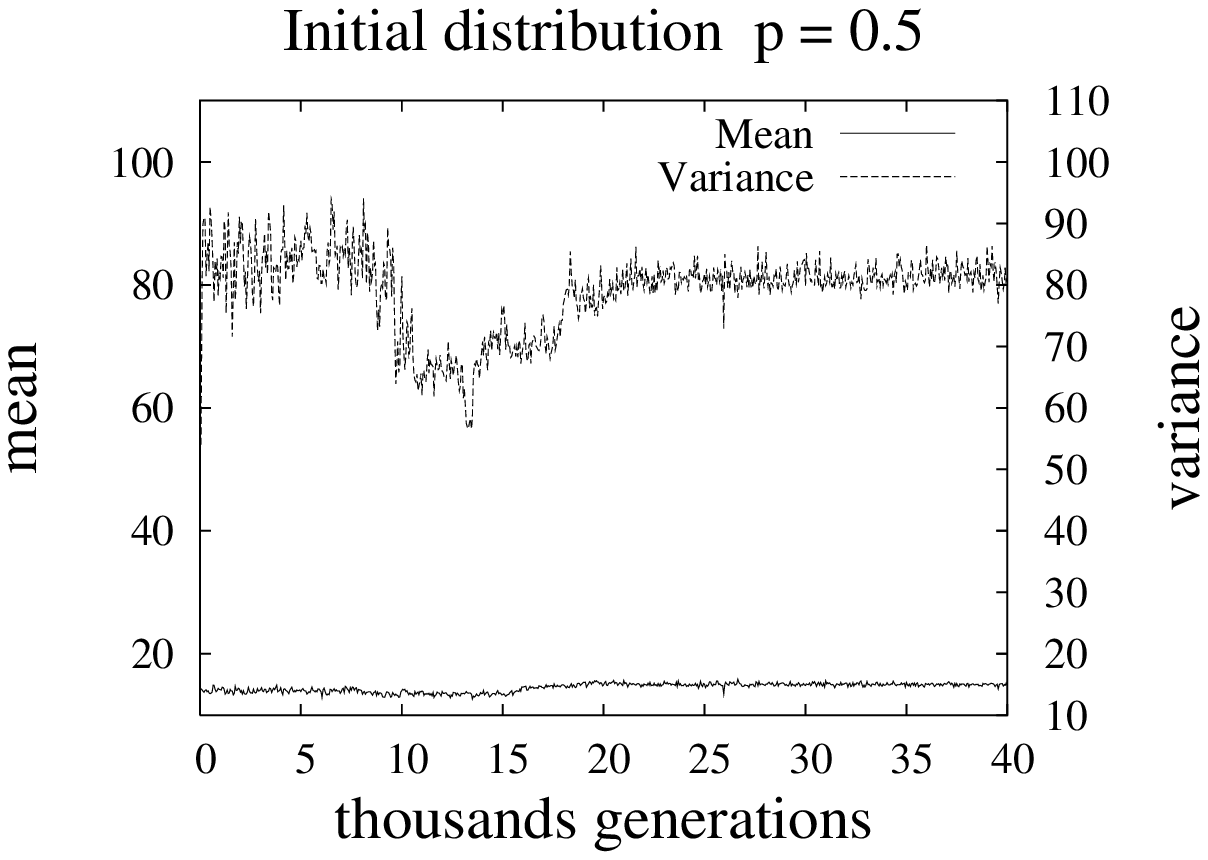}
\end{tabular}

\caption{Genome length $L = 28$. Plots of mean fitness, variance and mean obtained with  competition  intensity $J = 1$ and   assortativity $\Delta = 2$ in the case of flat fitness landscape ($\beta = 100$, $\Gamma = 28$) . Annealing parameters: $\mu_0 = (10)^{-1}$, $\mu_{\infty} = (10)^{-6}$, $\tau = 10000$, $\delta = 3000$. Total evolution time 40000 }

\label{fitness-variance_9}
\end{center}
\end{figure}

The evolutionary pattern for $L=28$ is basically the same as with $L=14$. The initial distribution quickly extends over the whole phenotype space with higher frequencies in the regions near $x=0$ and $x=28$. The phenotypes in the range $x=1$ to $x=5$ and $x=23$ to $x=27$ corresponds roughly to $5 \%$ of the population each, while the frequencies of phenotypes from $x=6$ to $x=22$ range from 2.5 to 3 $\%$. It therefore appears that the positions of the peaks when $L=28$ are roughly the same as the positions observed with $L=14$ multiplied by 2. The creation of this wide distribution corresponds to a significant increase in mean fitness and variance. After $T = \tau$ the intermediate phenotypes tend to disappear and the final distribution will be represented by two or three delta peaks. When the final distribution is represented by three delta peaks the first one is typically located in $x=2$, $x=3$, $x=4$ and less often also in $x=1$ and $x=5$; the second peak typically lies in the range 13 to 16 and the third peak in the range 23 to 26. When only two peaks appear in the stationary distribution, the first one is usually in the range 7 to 9 while the second one is in the range 21 to 24. The appearance of these delta peaks after $T = \tau$ is characterized by a significant decrease both in mean fitness and variance. The decrease in mean fitness is due to the increase in intraspecific competition that occurs when the population initially spread over the whole phenotype space concentrates on few phenotypes only. The decrease in variance conversely, is due to the extinction of the intermediate phenotypes that provided to the variance many small contributions that summed up to a considerable value. The intermediate phenotypes are replaced by a single peak not far from the mean of the distribution whose contribution to the variance is very small. The stationary distribution and the plots of mean fitness, variance and fitness are shown in Figures~\ref{fig:9} and~\ref{fitness-variance_9}.

\begin{figure}[ht!]
\begin{center}
\includegraphics[scale=0.7]{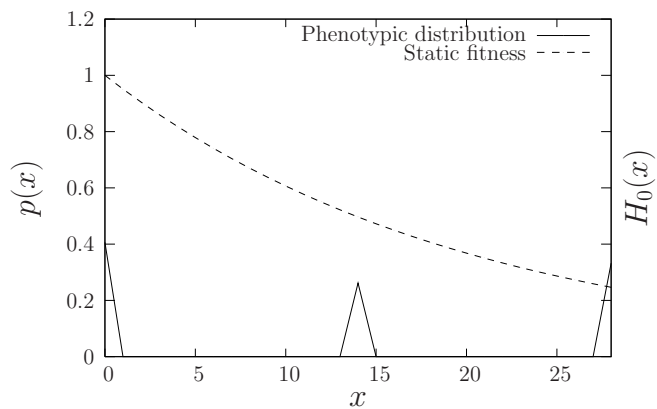}

\caption{Genome length $L = 28$.  The final distribution (generation 40000) obtained with  competition  intensity $J = 10$ and   assortativity $\Delta = 6$ in the case of steep fitness landscape ($\beta = 1$, $\Gamma = 20$) . Annealing parameters: $\mu_0 = (10)^{-1}$, $\mu_{\infty} = (10)^{-6}$, $\tau = 20000$, $\delta = 1000$. Total evolution time 40000 }
\label{fig:10}
\end{center}
\end{figure}

\begin{figure}[ht!]
\begin{center} 
\begin{tabular}{cc}
\includegraphics[scale=0.6]{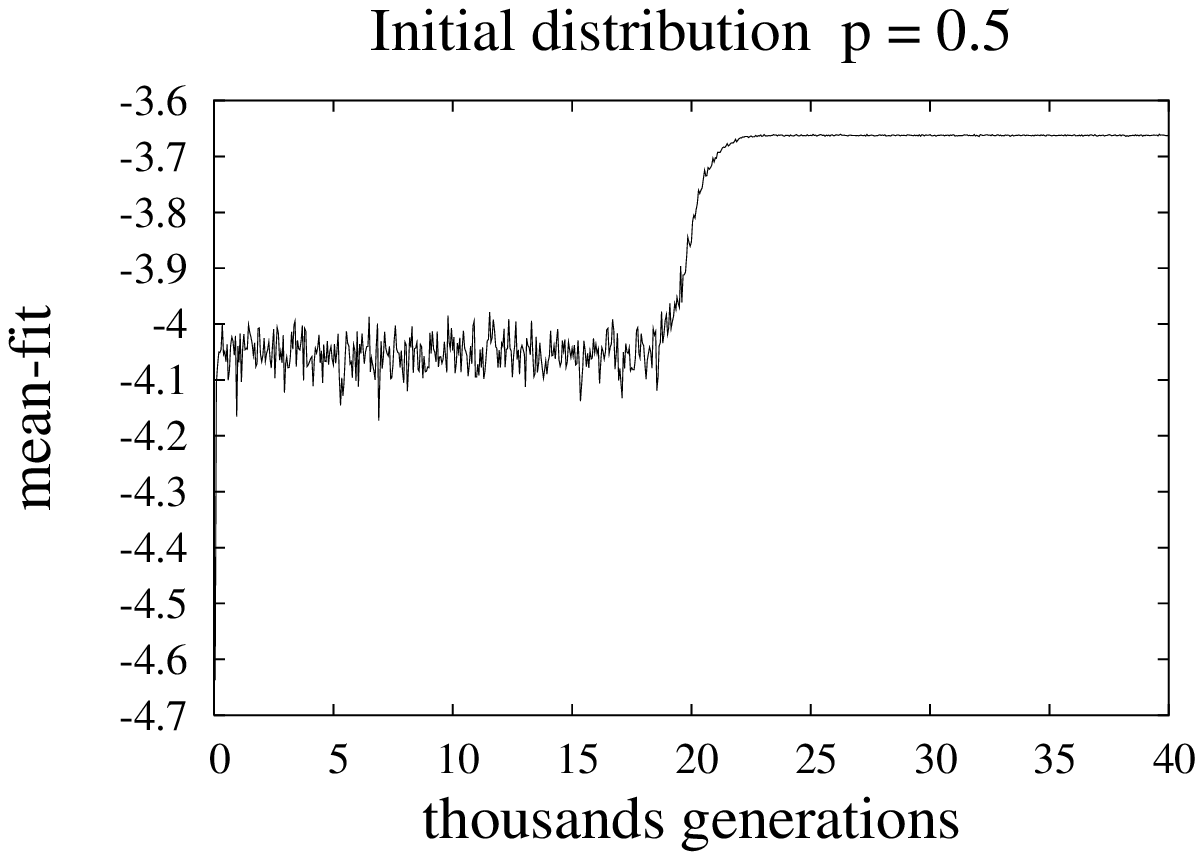} &
\includegraphics[scale=0.6]{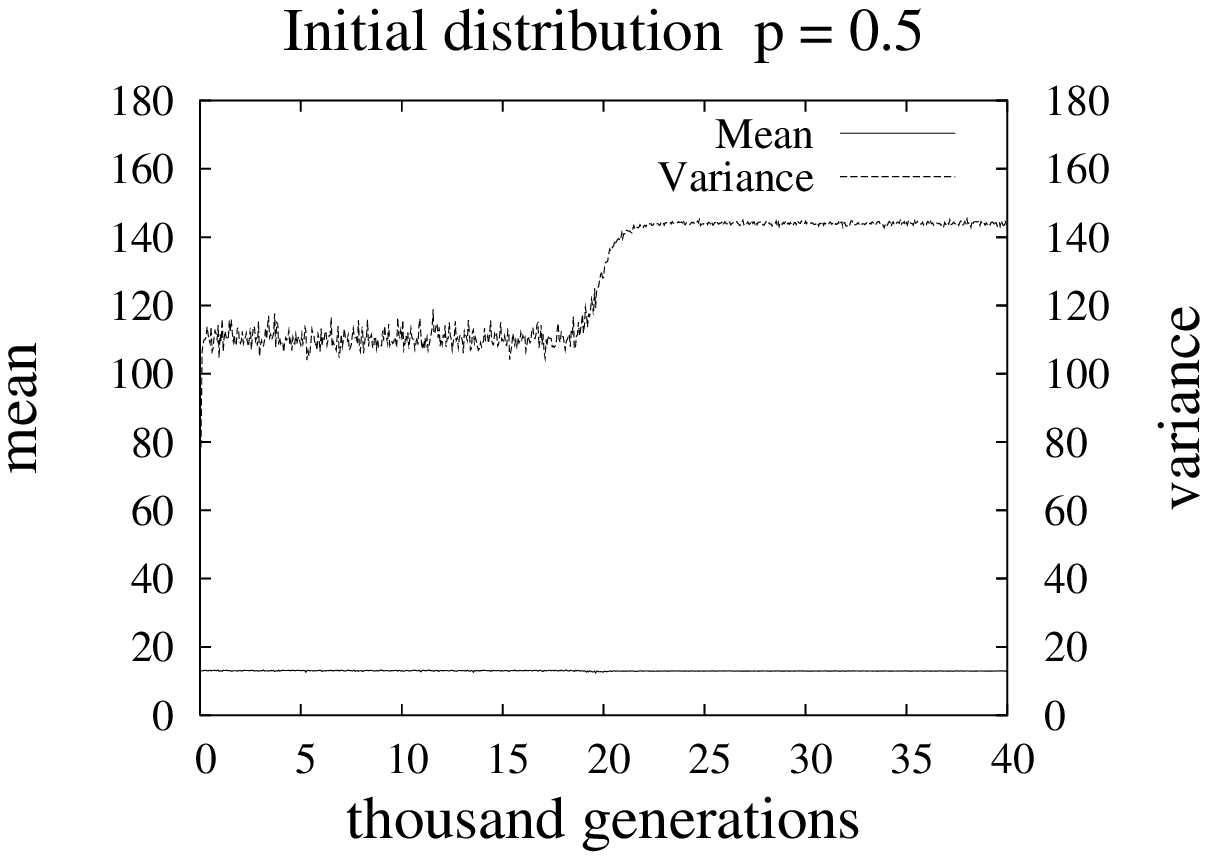}
\end{tabular}

\caption{Genome length $L = 28$. Plots of mean fitness, variance and mean obtained with  competition  intensity $J = 10$ and   assortativity $\Delta = 6$ in the case of steep fitness landscape ($\beta = 1$, $\Gamma = 20$) . Annealing parameters: $\mu_0 = (10)^{-1}$, $\mu_{\infty} = (10)^{-6}$, $\tau = 20000$, $\delta = 1000$. Total evolution time 40000 }

\label{fitness-variance_10}
\end{center}
\end{figure}

We now consider a simulation with a steep static fitness landscapes that can be compared with the one shown in Figures~\ref{fig:7B3} and~\ref{fitness-variance_7B3}. Also in this case the doubling of $L$ to 28 was paralleled by a doubling of $\Gamma$, $R$ and $\Delta$ to 20, 8 and 6 respectively. The initial distribution splits in three distributions at the opposite ends and in the middle of the phenotypic space, which leads to a substantial increase in mean fitness and variance. After $T=\tau$ the three distributions become delta peaks in $x=0$, $x=14$ (but occasionally also in $x=15$) and in $x=28$. When this happens a second increase in mean fitness and variance takes place. The stationary distribution and the plots of mean fitness, variance and fitness in a simulation with $p=0.5$ are shown in Figures~\ref{fig:10} and~\ref{fitness-variance_10}.

\newpage

\bibliographystyle{unsrt}

\bibliography{references_speciation}

\begin{thebibliography}{10}

\bibitem{ciclidi}
U.K. Schlieven, D.~Tautz, and S.~P{\"a}{\"a}bo.
\newblock Sympatric speciation suggested by monophyly of crater lake cichlids.
\newblock {\em Nature}, 368:629--632, 1994.

\bibitem{stickleback1}
D.~Schluter.
\newblock Experimental evidence that competition promotes divergence in
  adaptive radiation.
\newblock {\em Science}, 266:798--801, 1994.

\bibitem{stickleback2}
L.~Nagel and D.~Schluter.
\newblock Body size, natural selection, and speciation in sticklebacks.
\newblock {\em Evolution}, 52:209--218, 1998.

\bibitem{stickleback3}
H.D. Rundle and D.~Schluter.
\newblock Reinforcement of stickleback mate preferences: Sympatry breeds
  contempt.
\newblock {\em Evolution}, 52:200--208, 1998.

\bibitem{stickleback4}
E.B. Taylor and J.D. McPhail.
\newblock {\em Biological Journal of the Limnean Society}, 66:271--291, 1999.

\bibitem{snails}
K.~Johannesson, E.~Rolan-Alvarez, and A.~Ekendahl.
\newblock Incipient reproductive isolation between two sympatric morphs of the
  intertidal snail littorina saxatilis.
\newblock {\em Evolution}, 49:1180--1190, 1995.

\bibitem{lizards}
J.B. Losos, T.R. Jackman, A.~Larson, K.~de~Queiroz, and L.~Rodriguez-Schettino.
\newblock Contingency and determinism in replicated adaptive radiations of
  island lizards.
\newblock {\em Science}, 279:2115--2118, 1998.

\bibitem{Senecio}
E.B. Knox and J.D. Palmer.
\newblock Chloroplast dna variation and the recent radiation of the giant
  senecios (asteraceae) on the tall mountains of eastern africa.
\newblock {\em Proceedings of the National Academy of Sciences USA},
  92:10349--10353, 1995.

\bibitem{Travisano}
P.B. Rainey and M~Travisano.
\newblock Adaptive radiation in a heterogeneous environment.
\newblock {\em Nature}, 394:69--72, 1999.

\bibitem{KK}
A.S. Kondrashov and F.A. kondrashov.
\newblock Interactions amon quantitative traits in the course of sympatric
  speciation.
\newblock {\em Nature}, 400:351--354, 1999.

\bibitem{DD2}
U.~Dieckmann and M.~Doebeli.
\newblock Evolutionary branching and sympatric speciation caused by different
  types of ecological interactions.
\newblock IIASA Interim Report IR-00-040, July 2000.

\bibitem{DD1}
U.~Dieckmann and M.~Doebeli.
\newblock On the origin of species by sympatric speciation.
\newblock {\em Nature}, 400:354--357, 1999.

\bibitem{Price}
G.R. Price.
\newblock Fisher's fundamental theorem made clear.
\newblock {\em Annals of human genetics}, 36:129--140, 1972.

\bibitem{Ewens}
W.J. Ewens.
\newblock An interpratation and proof of the fundamental theorem of natural
  selection.
\newblock {\em Theoretical population biology}, 36:167--180, 1989.

\bibitem{Bagnoli}
F.~Bagnoli and M.~Bezzi.
\newblock An evolutionary model for simple ecosystems.
\newblock In Dietrich Stauffer, editor, {\em Annual Reviews of Computational
  Physics VII}, pages 265--309. World Scientific Publishing Company, 1999.

\bibitem{Bagnoli2}
F.~Bagnoli and M.~Bezzi.
\newblock {\em Physical Review Letters}, 79:3302, 1997.

\bibitem{FitnessLandscape}
D.~Sherrington.
\newblock Landscape paradigm in physics and biology.
\newblock {\em Physica D}, 107:117, 1997.

\bibitem{Peliti1}
L.~Peliti.
\newblock Fitness landscapes and evolution.
\newblock In T.~Riste and D.~Sherrington, editors, {\em Physics of
  Biomaterials: Fluctuations, Self-Assembly and Evolution}, pages 287--308.
  Dordrecht:Kluwer, 1996.

\bibitem{Peliti2}
L.~Peliti.
\newblock Introduction to the statistical theory of darwinian evolution.
\newblock Lectures given at ICTP Summer College on frustrated systems, Trieste.
  Notes taken by Ugo Bastolla and Susanna Manrubia, August 1997.

\bibitem{Fisher1}
R.A. Fisher.
\newblock {\em The genetical theory of natural selection}.
\newblock Clarendon Press, Oxford, 1930.

\bibitem{Fisher2}
R.A. Fisher.
\newblock Average excess and average effect of a gene substitution.
\newblock {\em Annals of Eugenics}, 11:53--63, 1941.

\bibitem{Fisher3}
R.A. Fisher.
\newblock {\em The genetical theory of natural selection}.
\newblock Dover Publications, New York, 2nd edition, 1958.

\bibitem{Holland}
J.H. Holland.
\newblock {\em Adaptation in natural and artificial systems}.
\newblock MIT Press, 1975.

\bibitem{BagnoliGuardianiRicombinazione}
F.~Bagnoli and C.~Guardiani.
\newblock A microscopic model of evolution of recombination.
\newblock {\em to appear in Physica A}, 2004.

\end{thebibliography}

\end{document}